\documentclass[a4paper,11pt]{article}
\usepackage[left=2.5cm,right=2.5cm]{geometry}
\usepackage{cite}
\usepackage{amsmath,amsfonts,amssymb}
\usepackage{graphicx}

\title{Study of the $\gamma d\to K^{+}K^{-}np$ reaction and an alternative explanation for the ``$\Theta^{+}(1540)$ pentaquark'' peak}
\author{A. Mart\'inez Torres\footnote{amartine@ific.uv.es}, $\,\,$E.~Oset\footnote{oset@ific.uv.es}\\\\
{\small{ \it Departamento de F\'isica Te\'orica and IFIC,
Centro Mixto Universidad de Valencia-CSIC,}}\\
{\small{\it Institutos de
Investigaci\'on de Paterna, Aptdo. 22085, 46071 Valencia, Spain.}}
}

\date{\today}

\begin{document}
\maketitle
\abstract {We present a calculation of the $\gamma ~d \to ~K^+ K^- ~n ~p $ reaction with the aim of seeing  if the experimental peak observed in the $K^+ n$ invariant mass around 1526 MeV, from where evidence for the existence of the $\Theta^+$ has been claimed, can be obtained without this resonance as a consequence of the particular dynamics of the process and the cuts applied in the experimental set up.  We find that a combination of facts leads indeed to a peak around 1530 MeV for the invariant mass of $K^+ n$ without the need to invoke any new resonance around this energy. This, together with statistical fluctuations that we prove to be large with the statistics of the experiment, is likely to produce the narrower peak observed there.}

\section{Introduction}
 The discovery of a peak in the $\gamma \, ^{12}C \to K^+ K^- X$ reaction in the
$K^+n$ invariant mass spectrum, which was identified as a signal for a
pentaquark of positive strangeness, the $\Theta^+$ \cite{nakaone}, created 
a turmoil in the Hadron Physics Community and stimulated many 
theoretical and experimental works. A large number of experiments were conducted,
many of which found also the signal, but gradually other experiments were
performed that did not produce the expected peak.  The number of theoretical
works stimulated by this finding in the following years was of the order of one thousand and, whether they could or could not find a justification for the 
$\Theta^+$ state, they gave collectively a big push towards the understanding of 
the hadronic spectra.  A comprehensive review of these developments was done
after some time in \cite{hicks}, were one can see the relevant literature on the
subject.  More recently a new experiment was done at LEPS on a deuteron target, 
and with more statistics, and a clear peak was observed around 1526 MeV in the
$K^+n$ invariant mass distribution \cite{nakatwo}. While quite some evidence is given for the existence of the peak in \cite{nakatwo}, it is not ruled out that it could be a consequence of the peculiar dynamics of the process, with the experimental set up and the cuts applied to see the peak, rather than the signal of a genuine new resonance.  The fact that the same reaction measured at CLASS with ten times more statistics and complete kinematics fails to see the $\Theta^{+}$ peak \cite{kinnon}, is certainly an incentive to search for some alternative explanation of the peak observed at LEPS. A first thought comes to our mind: in the reaction of \cite{nakatwo}, $\gamma ~d \to ~K^+ K^- ~n ~p $, the main contribution comes from  $\phi$ production, which is removed by particular cuts done in the $K^+ K^-$ invariant mass in \cite{nakatwo}. This, together with the fact that the particles are detected in the forward direction could lead to some unwanted structure in the mass distribution of $K^+ ~n$. It should not be overlooked the fact that in \cite{nakatwo} only the $K^+$ and the $ K^-$ are detected, but not the $p$ and the $n$. The reconstruction of the $K^+ ~n$ invariant mass is done using a prescription to eliminate as much as possible the effects of Fermi motion and to have the $p$ as a spectator. We shall show in the present paper that the prescription works to show peaks coming from genuine resonances, but also creates  a distorted spectrum.  On the other hand, the procedure does not work in the reverse direction, and if a peak appears it is not necessarily an evidence of the presence of a resonance. 

  There is another issue which requires attention. The experiment is done in deuterium, since one is interested in having the $K^+ ~n$ invariant mass. This means one necessarily will have rescattering of the produced kaons. Hence, assuming that a $\phi$ is created and one of the kaons rescatters from a second nucleon, the rescattered kaon changes energy and direction and the reconstructed $K^+ K^-$ invariant mass does no longer match the original $\phi$ from where the two kaons came from. Since the $\phi$ production counts for the largest part of the cross section in the experiment of \cite{nakatwo}, it could be that this part of the amplitude accounts for a relatively sizable part of the total amplitude and through interference with the one body processes leads to a some kind of signal which might be misidentified in the analysis. 

      There are, thus, some reasons that make us think that some peak could be produced, not tied to the existence of a resonance but to this peculiar combination of dynamics, multiple scattering and cuts. The present paper presents a thorough simulation of the physical process, using a theoretical model that accounts for $\phi$ production and rescattering of the kaons. The production of the $\Lambda(1520)$ is also considered and the procedure to filter events with a nucleon as spectator is tested with this resonance. The rescattering part versus the one body part of the amplitude is done accurately, using realistic deuteron wave functions and amplitudes for $K N \to KN$ and 
$\bar{K} N \to \bar{K}N$ constructed using chiral unitary dynamics. 

     The results obtained show that a peak around 1530 MeV
for the invariant mass of the $K^+ n$ pair appears, even with large statistics, without having any $\Theta^+$ term in the amplitude. The peak obtained is broad. However, with the statistics of the experiment of \cite{nakatwo}, fluctuations appear often which can produce a narrower peak in that region, very similar to that obtained in \cite{nakatwo}. We also show that it is possible, after some strength appears in that region with a predetermined set up, to obtain a sharper peak by changing a bit this set up.  All these findings give us strong reasons to think that the peak seen in \cite{nakatwo} obeys to this combination of facts related to the experimental set up and is not a signal for a new resonance.

  The work proceeds as follows.   In section \ref{forma} we explain the formalism to calculate the cross section for the process $\gamma d\to K^{+}K^{-}np$ with detail and the model to take into account the $\phi$ and $\Lambda(1520)$ productions. In section \ref{penphi} we show the $K^{+}n$ invariant mass distribution obtained with the same set up than at LEPS . In section \ref{in1520} we show the results that we obtain when we introduce the $\Lambda(1520)$, in section \ref{resca} the rescattering contribution and in section \ref{sta1} we make a statistical analysis and compare with the results of LEPS, after which we draw our conclusions.

\section{Formalism}\label{forma}
When trying to simulate the $\gamma d\to K^{+}K^{-}np$ reaction one must consider the basic features learned from the experiment in \cite{nakatwo}. The most important contribution comes from $\phi$ production, the elementary reaction $\gamma p\to \phi p$ and $\gamma n\to \phi n$. We will implement this $\phi$ production in our simulation through a minimal model which incorporates the basic structure of $K^{+}K^{-}$ production correlated by the $\phi$ propagator. We will also take into account empirical information from \cite{nakaphi}, from where one induces similar $\phi$ cross section production from proton or neutron targets. 

Another element in the dynamics of the $\gamma d\to K^{+}K^{-}np$ reaction is the production of the $\Lambda(1520)$. Although its contribution to the process is small compared to that of $\phi$ production, we will take advantage of it to test the method used in \cite{nakatwo} to eliminate Fermi motion in the reaction on deuteron.

An important ingredient in our approach is the rescattering that unavoidably occurs in the reaction in deuteron. Indeed, consider a first step of $\gamma p\to \phi p\to K^{+}K^{-}p$ and then a rescattering of the $K^{+}$ or the $K^{-}$ with the neutron. This double scattering step can be evaluated theoretically, and its ratio to the first step $\gamma p\to\phi p$ can be calculated rather reliably. This is so since uncertainties in the elementary $\phi$ production model cancel in the ratio and the rest roughly requires the knowledge of the deuteron wave function and the scattering amplitude of $K^{+}$ ($K^{-}$) with the nucleons, which one has rather under control from studies of the $\bar{K}N$ and $KN$ interaction using unitary chiral dynamics \cite{ramos,bennhold,meissner,meissner2,magas,felipe}.

\begin{figure}[t!]
\includegraphics[width=\textwidth]{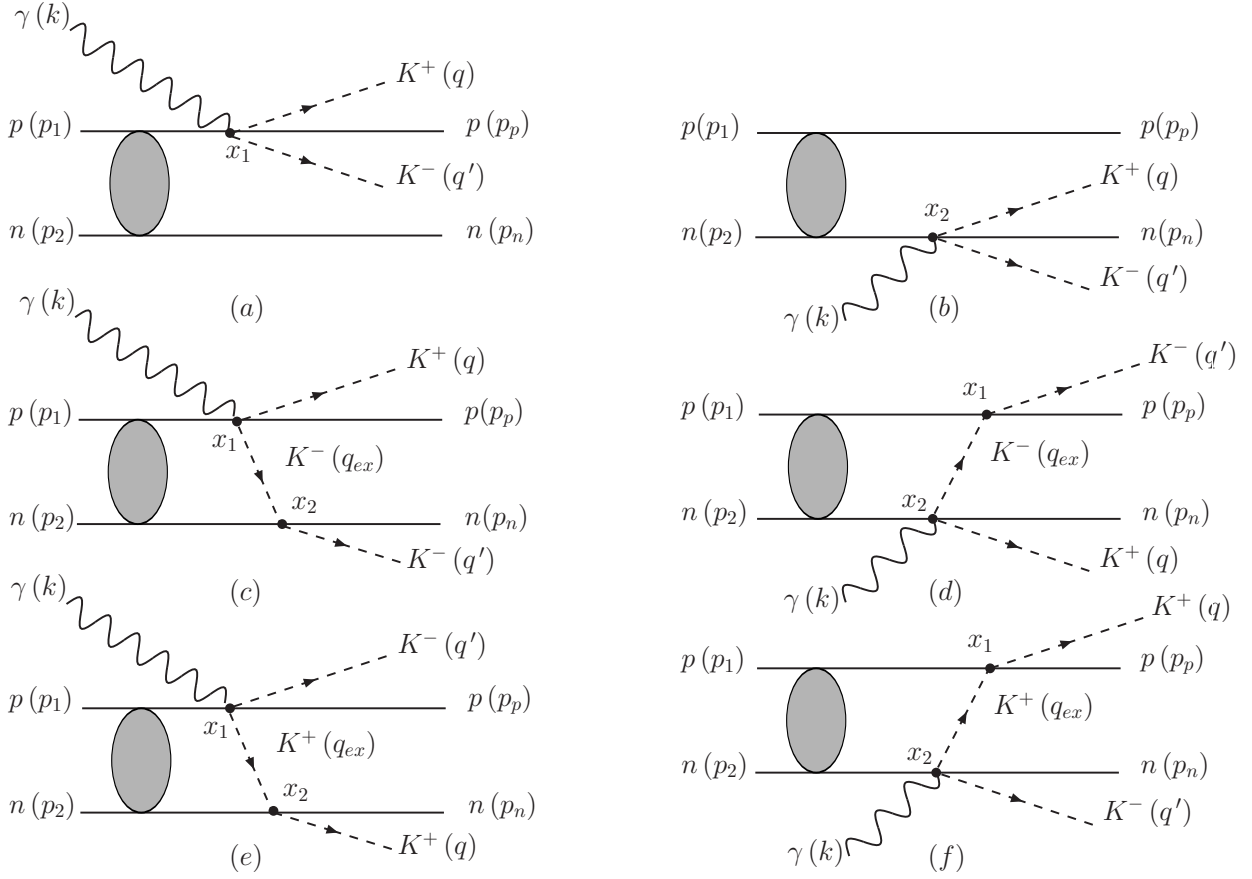}
\caption{(Color online) Diagrammatic representation of the model for the $\gamma d\to K^{+}K^{-}np$ reaction.}\label{model}
\end{figure}

The consideration of the rescattering is important in principle in this simulation. Indeed, the $\phi$ production takes the largest strength of the $\gamma p\to K^{+} K^{-}p$ reaction in the energy region of LEPS. Yet, with the cuts of LEPS one removes this contribution, that would show up as a large background on the $K^{+}n$ spectrum precluding the observation of some smaller signal. However, if one has $\phi$ production and rescattering of one of the kaons, this latter kaon will change energy and direction upon rescattering, such that when the cut on the $K^{+}K^{-}$ invariant mass is implemented, these events are not removed. The kinematics of these events coming from a primary $\phi$ production is very peculiar, such that it is not unlikely that they could lead to some peak in the $K^{+}n$ invariant mass distribution which could be misidentified as a signal of some resonance.

The model that we take is depicted diagrammatically in Fig. \ref{model}. We have there $\phi$ production on the $p$ plus the same with rescattering of the $K^{+}$ or the $K^{-}$. Similarly we also have $\phi$ production on the neutron plus the same with rescattering of the $K^{+}$ or the $K^{-}$. In the case of $\gamma p\to K^{+}K^{-}p$ we also can have the $K^{-}p$ making the $\Lambda(1520)$, which we will also take into account. The technical details follow in the next subsections.

\subsection{The impulse approximation}
In order to calculate the cross section for the process $\gamma d\to K^{+}K^{-}np$ we need to determine the $S$ matrix or $T$ matrix for this reaction. Let us consider the first diagram in Fig. \ref{model}, corresponding to the impulse approximation, and let us evaluate its contribution to the $S$ matrix.
Considering plane waves for the wave functions of the incident photon and the particles in the final state, the $S$-matrix for the diagram of Fig. \ref{model}a can be expressed as
\begin{align}
S^{(a)}&=\int d^{4}x_{1}\frac{1}{(\sqrt{V})^{4}}\mathcal{N}_{p}\mathcal{N}_{K^{+}}\mathcal{N}_{K^{-}}\mathcal{N}_{\gamma}e^{i(p_{p}+q+q^{\prime}-k)x_{1}}\nonumber\\
&\quad\times(-it^{(a)}_{\gamma p\to K^{+}K^{-}p})\mathcal{N}_{1} e^{-ip^{0}_{1}x^{0}_{1}}\varphi_{1}(\vec{x}_{1})\int d^{3}x_{2}\frac{1}{\sqrt{V}}\mathcal{N}_{n}e^{-i\vec{p}_{n}\vec{x}_{2}}\mathcal{N}_{2}\varphi_{2}(\vec{x}_{2})\label{Sa_a}
\end{align}
 where $\varphi_{i}$ ($i=1,2$) are the wave functions of the nucleons in the deuteron, $p^{0}_{1}$ is the energy of the proton inside the deuteron, $p_{p}$, $p_{n}$, $q$, $q^{\prime}$ correspond to the four momentum of the proton, neutron, $K^{+}$ and $K^{-}$ of the final state, respectively, and $k$ to the four momentum of the incident photon. The normalization factor $\mathcal{N}_{i}$ is given by $\mathcal{N}_{i}=\sqrt{M_{i}/E_{i}}$ for baryons and $\mathcal{N}_{i}=1/\sqrt{2\omega_{i}}$ for mesons. The plane waves are normalized inside a box with a volume $V$.
 
 The calculation of $S^{(a)}$ requires the evaluation of the integrals in the $x_{1}$ and $x_{2}$ variables.
In order to perform the integration in $x_{1}$ we separate the temporal and the spatial components and write
\begin{align}
\int d^{4}x_{1}e^{i(p_{p}+q+q^{\prime}-k)x_{1}}e^{-ip^{0}_{1}x^{0}_{1}}=\int dx^{0}_{1}e^{i(p^{0}_{p}+q^{0}+{q^{\prime}}^{0}-k^{0}-p^{0}_{1})x^{0}_{1}}\int d^{3}x_{1}e^{-i(\vec{p}_{p}+\vec{q}+\vec{q}^{\,\prime}-\vec{k})\vec{x}_{1}}.\nonumber
\end{align}
The integration over the time component $x^{0}_{1}$ gives the delta function for the energy conservation
\begin{align}
\int dx^{0}_{1}e^{i(p^{0}_{p}+q^{0}+{q^{\prime}}^{0}-k^{0}-p^{0}_{1})x^{0}_{1}}=2\pi\delta(k^{0}+p^{0}_{1}-p^{0}_{p}-q^{0}-{q^{\prime}}^{0}).\label{del}
\end{align}
In the impulse approximation the energy of the spectator, in this case the neutron, does not change, i.e., $p^{0}_{n}=p^{0}_{2}$, with $p^{0}_{2}$ the energy of the neutron inside the deuteron (omitting the small binding energy). Therefore, we can write the $\delta$ in Eq. (\ref{del}) as
\begin{align}
\delta(k^{0}+p^{0}_{1}-p^{0}_{p}-q^{0}-{q^{\prime}}^{0})=\delta(k^{0}+p^{0}_{1}+p^{0}_{2}-p^{0}_{p}-p^{0}_{n}-q^{0}-{q^{\prime}}^{0}).
\end{align}

The evaluation of the spatial integrals is done by making the usual transformation
\begin{align}
\vec{R}&=\frac{1}{2}(\vec{x}_{1}+\vec{x}_{2})\nonumber\\
\vec{r}&=\vec{x}_{1}-\vec{x}_{2}\label{CM}
\end{align}
with $\vec{R}$ being the center of mass position of the $p n$ system (assuming an average mass for the proton and the neutron) and $\vec{r}$ the
relative coordinate between $\vec{x}_{1}$ and $\vec{x}_{2}$. In this way, the term

\begin{align}
\int d^{3}x_{1}\int d^{3}x_{2}e^{-i(\vec{p}_{p}+\vec{q}+\vec{q}^{\,\prime}-\vec{k})\vec{x}_{1}}e^{-i\vec{p}_{n}\vec{x}_{2}}\mathcal{N}_{1}\mathcal{N}_{2}\varphi_{1}(\vec{x}_{1})\varphi_{2}(\vec{x}_{2})
\end{align}
present in Eq. (\ref{Sa_a}) can be written as
\begin{align}
\int d^{3}R \,e^{-i(\vec{p}_{p}+\vec{q}+\vec{q}^{\,\prime}-\vec{k}-\vec{p}_{d}+\vec{p}_{n})\vec{R}}\int d^{3}r e^{-\frac{i}{2}(\vec{p}_{p}+\vec{q}+\vec{q}^{\,\prime}-\vec{k}-\vec{p}_{n})\vec{r}}\frac{1}{\sqrt{V}}\mathcal{N}_{d}\varphi(\vec{r})\label{intr}
\end{align}
where we have introduce the deuteron wave function $\varphi(\vec{r})$ for the relative motion
\begin{align}
\mathcal{N}_{1}\mathcal{N}_{2}\varphi_{1}(\vec{x}_{1})\varphi_{2}(\vec{x}_{2})\to\frac{1}{\sqrt{V}}\mathcal{N}_{d}e^{i\vec{p}_{d}\vec{R}}\varphi(\vec{r})
\end{align}
with $\vec{p}_{d}$ the momentum of the deuteron. The deuteron wave function is normalized as
\begin{align}
\int d^{3}r |\varphi(\vec{r})|^{2}=1.
\end{align}
 We only consider the s-wave component of the deuteron wave function, which in the rest frame of the deuteron can be parametrized as \cite{deuteron}
 \begin{align}
 \varphi(\vec{r})=\sum_{i=1}^{11}\frac{C_{i}}{r}exp(-m_{i}r).
\end{align}
The coefficients $C_{i}$ and $m_{i}$ can be found in \cite{deuteron2}. The integration over $\vec{r}$ in Eq. (\ref{intr}) results into the Fourier transformation of the deuteron wave function, while the integration over $\vec{R}$ generates the delta function of the total momentum conservation

\begin{align}
\int d^{3}r\,e^{\frac{i}{2}(\vec{k}+\vec{p}_{n}-\vec{p}_{p}-\vec{q}-\vec{q}^{\,\prime})\vec{r}}\varphi(\vec{r})=\tilde{\varphi}\Bigg(\frac{\vec{k}+\vec{p}_{n}-\vec{p}_{p}-\vec{q}-\vec{q}^{\,\prime}}{2}\Bigg)\nonumber\\
\int d^{3}R\,e^{-i(\vec{p}_{p}+\vec{q}+\vec{q}^{\,\prime}-\vec{k}-\vec{p}_{d}+\vec{p}_{n})\vec{R}}=(2\pi)^{3}\delta^{3}(\vec{p}_{p}+\vec{q}+\vec{q}^{\,\prime}-\vec{k}-\vec{p}_{d}+\vec{p}_{n}).
\end{align}
Taking the deuteron energy as the sum of the energies of the nucleons, i.e., $p^{0}_{d}=p^{0}_{1}+p^{0}_{2}$,  we have, collecting the previous results,
\begin{align}
S^{(a)}&=-i\frac{1}{V^{3}}\mathcal{N}_{d}\mathcal{N}_{\gamma}\mathcal{N}_{p}\mathcal{N}_{n}\mathcal{N}_{K^{+}}\mathcal{N}_{K^{-}}(2\pi)^{4}\delta(p^{0}_{d}+k^{0}-p^{0}_{p}-p^{0}_{n}-q^{0}-{q^{0}}^{\prime})\nonumber\\
&\quad\times\delta^{3}(\vec{p}_{p}+\vec{p}_{n}+\vec{q}+\vec{q}^{\,\prime}-\vec{k}-\vec{p}_{d})\tilde{\varphi}\Bigg(\frac{\vec{k}+\vec{p}_{n}-\vec{p}_{p}-\vec{q}-\vec{q}^{\,\prime}}{2}\Bigg)\label{Sa_b}
\nonumber\\
&\quad\times t^{(a)}_{\gamma p\to K^{+}K^{-}p}.\end{align}
Using the momentum conservation law for the reaction $\gamma d\to K^{+}K^{-}np$
\begin{align}
\vec{k}+\vec{p}_{d}=\vec{q}+\vec{q}^{\,\prime}+\vec{p}_{p}+\vec{p}_{n}\label{conser}
\end{align}
the argument of the $\tilde{\varphi}$ function in Eq. (\ref{Sa_b}) can be put as
\begin{align}
\tilde{\varphi}\Bigg(\frac{\vec{k}+\vec{p}_{n}-\vec{p}_{p}-\vec{q}-\vec{q}^{\,\prime}}{2}\Bigg)=\tilde{\varphi}\Bigg(\vec{p}_{n}-\frac{\vec{p}_{d}}{2}\Bigg).
\end{align}
and, therefore, Eq. (\ref{Sa_b}) reads as
\begin{align}
S^{(a)}&=-i\frac{1}{V^{3}}\mathcal{N}_{d}\mathcal{N}_{\gamma}\mathcal{N}_{p}\mathcal{N}_{n}\mathcal{N}_{K^{+}}\mathcal{N}_{K^{-}}(2\pi)^{4}\delta^{4}(k+p_{d}-p_{p}-p_{n}-q-q^{\,\prime})\nonumber\\
&\quad\times\tilde{\varphi}\Bigg(\vec{p}_{n}-\frac{\vec{p}_{d}}{2}\Bigg)
 t^{(a)}_{\gamma p\to K^{+}K^{-}p}
 \end{align}
 Calling
 \begin{align}
 \mathcal{A}=\frac{1}{V^{3}}\Bigg(\prod_{i}\mathcal{N}_{i}\Bigg)(2\pi)^{4}\delta^{4}(k+p_{d}-p_{p}-p_{n}-q-q^{\,\prime})\label{A}
\end{align}
we arrive to
\begin{align}
S^{(a)}=-i \mathcal{A}\tilde{\varphi}\Bigg(\vec{p}_{n}-\frac{\vec{p}_{d}}{2}\Bigg)
 t^{(a)}_{\gamma p\to K^{+}K^{-}p}.
\end{align}
Similarly, it is straight forward to show that
\begin{align}
S^{(b)}&=-i\mathcal{A}\tilde{\varphi}\Bigg(\frac{\vec{p}_{d}}{2}-\vec{p}_{p}\Bigg)
 t^{(b)}_{\gamma n\to K^{+}K^{-}n}.
\end{align}
\subsection{The rescattering terms}
Let us consider now the diagrams where a kaon rescatters with one of the nucleons of the deuteron as shown in Figs. \ref{model}c-\ref{model}f. Analogously to the single scattering diagrams evaluated in the previous section, the $S$ matrix contribution for the diagram Fig. \ref{model}c is given by
\begin{align}
S^{(c)}&=\int d^{4}x_{1}\int d^{4}x_{2}\Bigg(\prod_{i}\mathcal{N}_{i}\Bigg)\frac{1}{(\sqrt{V})^{5}}e^{iq x_{1}}e^{iq^{\prime}x_{2}}e^{ip_{p}x_{1}}e^{ip_{n}x_{2}}\nonumber\\
&\quad\times\int\frac{d^{4}q_{ex}}{(2\pi)^{4}}\frac{i e^{iq_{ex}(x_{1}-x_{2})}}{q^{2}_{ex}-m_{K}^{2}+i\epsilon}e^{-ikx_{1}}e^{-ip^{0}_{1}x^{0}_{1}}\varphi_{1}(\vec{x}_{1})e^{-ip^{0}_{2}x^{0}_{2}}\varphi_{2}(\vec{x}_{2})\nonumber\\
&\quad\times(-i t^{(c)}_{\gamma p\to K^{+}K^{-}p})(-i t^{(c)}_{K^{-}n\to K^{-}n})\label{Sc}
\end{align}
Separating the temporal and the spatial part in the integration over the variables $x_{1}$ and $x_{2}$, we have, for the temporal part, \begin{align}
\int d x^{0}_{1}e^{i(q^{0}+p^{0}_{p}+q^{0}_{ex}-k^{0}-p^{0}_{1})x^{0}_{1}}&=(2\pi)\delta(k^{0}+p^{0}_{1}-q^{0}-p^{0}_{p}-q^{0}_{ex})\label{exp1}\\
\int d x^{0}_{2}e^{i({q^{\prime}}^{0}+p^{0}_{n}-q^{0}_{ex}-p^{0}_{2})x^{0}_{2}}&=(2\pi)\delta(q^{0}_{ex}+p^{0}_{2}-{q^{\prime}}^{0}-p^{0}_{n})\label{exp2}
\end{align}
Considering  $p^{0}_{d}=p^{0}_{1}+p^{0}_{2}$
and using Eqs. (\ref{exp1}), (\ref{exp2}), Eq. (\ref{Sc}) can be written as
\begin{align}
S^{(c)}&=-i\Bigg(\prod_{i}\mathcal{N}_{i}\Bigg)\frac{1}{(\sqrt{V})^{5}}\int\frac{d^{3}q_{ex}}{(2\pi)^{2}}\delta(k^{0}+p^{0}_{d}-q^{0}-{q^{\prime}}^{0}-p^{0}_{p}-p^{0}_{n})\nonumber\\
&\quad\times\int d^{3}x_{1}e^{-i(\vec{q}+\vec{p}_{p}+\vec{q}_{ex}-\vec{k})\vec{x}_{1}}\varphi_{1}(\vec{x}_{1})\int d^{3}x_{2}e^{-i(\vec{q}^{\,\prime}+\vec{p}_{n}-\vec{q}_{ex})\vec{x}_{2}}\varphi_{2}(\vec{x}_{2})\nonumber\\
&\quad\times t^{(c)}_{\gamma p\to K^{+}K^{-}p}t^{(c)}_{K^{-}n\to K^{-}n}\frac{1}{(q^{0}_{ex})^{2}-(\vec{q}_{ex})^{2}-m^{2}_{K}+i\epsilon}\Bigg |_{q^{0}_{ex}=k^{0}+p^{0}_{1}-q^{0}-p^{0}_{p}}\label{Sc_a}
\end{align}
where we have made use of Eq. (\ref{exp1}) in order to determine the value for $q^{0}_{ex}$.
Writing Eq. (\ref{Sc_a}) as a function of the $\vec{R}$ and $\vec{r}$ coordinates defined in Eq. (\ref{CM}) we have
\begin{align}
S^{(c)}&=-i\Bigg(\prod_{j}\mathcal{N}_{j}\Bigg)\frac{1}{(\sqrt{V})^{6}}\int \frac{d^{3}q_{ex}}{(2\pi)^{2}}\delta(k^{0}+p^{0}_{d}-q^{0}-{q^{\prime}}^{0}-p^{0}_{p}-p^{0}_{n})\nonumber\\
&\quad\times\int d^{3}R \,e^{-i(\vec{q}+\vec{q}^{\,\prime}+\vec{p}_{p}+\vec{p}_{n}-\vec{k}-\vec{p}_{d})\vec{R}}\int d^{3}r e^{-\frac{i}{2}(\vec{q}+\vec{p}_{p}+2\vec{q}_{ex}-\vec{k}-\vec{q}^{\,\prime}-\vec{p}_{n})\vec{r}}\varphi(\vec{r})\nonumber\\
&\quad\times t^{(c)}_{\gamma p\to K^{+}K^{-}p}t^{(c)}_{K^{-}n\to K^{-}n}\frac{1}{(q^{0}_{ex})^{2}-(\vec{q}_{ex})^{2}-m^{2}_{K}+i\epsilon}\Bigg |_{q^{0}_{ex}=k^{0}+p^{0}_{1}-q^{0}-p^{0}_{p}}\label{Sc_b}
\end{align}
with $j$ an index running over the final $p$, $n$, $K^{+}$, $K^{-}$ and the initial photon and deuteron.
The integration over $\vec{R}$ generates a delta function for the momentum conservation and the integration over $\vec{r}$ the corresponding Fourier transformation of the deuteron wave function $\varphi(\vec{r})$.  Taking into account this fact, Eq. (\ref{Sc_b}) can be expressed as
\begin{align}
S^{(c)}&=-i\Bigg(\prod_{j}\mathcal{N}_{j}\Bigg)\frac{1}{(\sqrt{V})^{6}}(2\pi)^{4}\delta^{4}(k+p_{d}-p_{p}-p_{n}-q-q^{\,\prime})\nonumber\\
&\quad\times\int\frac{d^{3}q_{ex}}{(2\pi)^{3}}\tilde{\varphi}\Bigg(\vec{q}^{\,\prime}+\vec{p}_{n}-\frac{\vec{p}_{d}}{2}-\vec{q}_{ex}\Bigg)
t^{(c)}_{\gamma p\to K^{+}K^{-}p}t^{(c)}_{K^{-}n\to K^{-}n}\nonumber\\
&\quad\times\frac{1}{(q^{0}_{ex})^{2}-(\vec{q}_{ex})^{2}-m^{2}_{K}+i\epsilon}\Bigg |_{q^{0}_{ex}=k^{0}+p^{0}_{1}-q^{0}-p^{0}_{p}}\label{Sc_c}
\end{align}
where we have used Eq. (\ref{conser}) for the argument of $\tilde{\varphi}$. In terms of Eq. (\ref{A}),  Eq. (\ref{Sc_c}) reads like
\begin{align}
S^{(c)}&=-i\mathcal{A}\int\frac{d^{3}q_{ex}}{(2\pi)^{3}}\tilde{\varphi}\Bigg(\vec{q}^{\,\prime}+\vec{p}_{n}-\frac{\vec{p}_{d}}{2}-\vec{q}_{ex}\Bigg)
t^{(c)}_{\gamma p\to K^{+}K^{-}p}t^{(c)}_{K^{-}n\to K^{-}n}\nonumber\\
&\quad\times\frac{1}{(q^{0}_{ex})^{2}-(\vec{q}_{ex})^{2}-m^{2}_{K}+i\epsilon}\Bigg |_{q^{0}_{ex}=k^{0}+p^{0}_{1}-q^{0}-p^{0}_{p}}.\label{Sc_d}
\end{align}

Proceeding in a similar way, it is possible to find the contribution for the remaining diagrams of Fig. \ref{model}, which we just write bellow
\begin{align}
S^{(d)}&=-i\mathcal{A}\int\frac{d^{3}q_{ex}}{(2\pi)^{3}}\tilde{\varphi}\Bigg(-\vec{q}^{\,\prime}-\vec{p}_{p}+\frac{\vec{p}_{d}}{2}+\vec{q}_{ex}\Bigg),
t^{(d)}_{\gamma n\to K^{+}K^{-}n}t^{(d)}_{K^{-}p\to K^{-}p}\nonumber\\
&\quad\times\frac{1}{(q^{0}_{ex})^{2}-(\vec{q}_{ex})^{2}-m^{2}_{K}+i\epsilon}\Bigg |_{q^{0}_{ex}=k^{0}+p^{0}_{2}-q^{0}-p^{0}_{n}}\label{Sd}\\
S^{(e)}&=-i\mathcal{A}\int\frac{d^{3}q_{ex}}{(2\pi)^{3}}\tilde{\varphi}\Bigg(\vec{q}+\vec{p}_{n}-\frac{\vec{p}_{d}}{2}-\vec{q}_{ex}\Bigg)
t^{(e)}_{\gamma p\to K^{+}K^{-}p}t^{(e)}_{K^{+}n\to K^{+}n}\nonumber\\
&\quad\times\frac{1}{(q^{0}_{ex})^{2}-(\vec{q}_{ex})^{2}-m^{2}_{K}+i\epsilon}\Bigg |_{q^{0}_{ex}=k^{0}+p^{0}_{1}-{q^{\,\prime}}^{0}-p^{0}_{p}},\label{Se}\\
S^{(f)}&=-i\mathcal{A}\int\frac{d^{3}q_{ex}}{(2\pi)^{3}}\tilde{\varphi}\Bigg(-\vec{q}-\vec{p}_{p}+\frac{\vec{p}_{d}}{2}+\vec{q}_{ex}\Bigg)
t^{(f)}_{\gamma n\to K^{+}K^{-}n}t^{(f)}_{K^{+}p\to K^{+}p}\nonumber\\
&\quad\times\frac{1}{(q^{0}_{ex})^{2}-(\vec{q}_{ex})^{2}-m^{2}_{K}+i\epsilon}\Bigg |_{q^{0}_{ex}=k^{0}+p^{0}_{2}-{q^{\,\prime}}^{0}-p^{0}_{n}}.\label{Sf}
\end{align}

The total $S$ matrix will be given by the sum of the contribution of the different diagrams
\begin{align}
S=1+\sum_{i=a}^{f}S_{i}
\end{align}
and the total T matrix is obtained using the relationship
\begin{align}
S=1-i\mathcal{A}\,T.\label{ST}
\end{align}

\subsection{Basic model for $\phi$ production and the $\bar{K}N$, $KN$ amplitudes}
\begin{figure}[h!]
\centering
\includegraphics[width=0.5\textwidth]{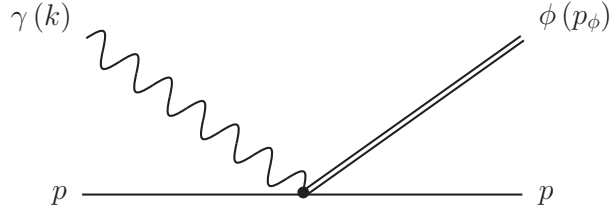}
\caption{Diagrammatic representation for the process $\gamma p\to \phi p$.}\label{phiprod}
\end{figure}

As shown in the former subsections, the calculation of the $S$ matrix involves the evaluation of the amplitudes for the processes $\gamma N \to \phi N\to K^{+}K^{-} N$, where N represents a nucleon, i.e., $N=$ $n$, $p$, and $\bar{K} N\to \bar{K}N$, $K N\to KN$.
In order to determine the amplitude for the process $\gamma N\to \phi N\to K^{+}K^{-}N$,  we are going to assume a minimal structure which satisfies gauge invariance. Let us first consider the process $\gamma N\to \phi N$ (Fig. \ref{phiprod}) . 

The amplitude which describes this process  must be a function depending on the four momentum of the photon, $k$, and the $\phi$, $p_{\phi}$, and their respective polarization vectors, $\epsilon_{\gamma}$ and $\epsilon_{\phi}$.  Imposing gauge invariance, this amplitude must be of the form
\begin{align}
t_{\gamma N\to \phi N}\sim(\epsilon_{\gamma}\epsilon_{\phi}k\cdot p_{\phi}-p_{\phi}\epsilon_{\gamma} k\epsilon_{\phi}).\label{coulomb}
\end{align}
If we work in the Coulomb gauge for the photons, $p_{\phi}\epsilon_{\gamma}=-\vec{p}_{\phi}\vec{\epsilon}_{\gamma}$, and considering $k\cdot p_{\phi}\sim k^{0} M_{\phi}$, we can neglect the second term in Eq. (\ref{coulomb}) as far as $|\vec{p}_{\phi}|<< M_{\phi}$. The particular structure of this amplitude is not essential in order to see the effects that the cuts of \cite{nakatwo} can produce in the $M_{K^{+}n}$ invariant mass distribution. Since we will evaluate the cross section in a reduced range of the photon energy, $k^{0}$, we can further assume this to be constant and thus \begin{align}
t_{\gamma N\to \phi N}=\tilde{a}_{N}\epsilon_{\gamma}\epsilon_{\phi}
\end{align}
where $\tilde{a}_{N}$ is considered as a parameter which determine the magnitude of the amplitude for this process. 
\begin{figure}[h!]
\centering
\includegraphics[width=0.7\textwidth]{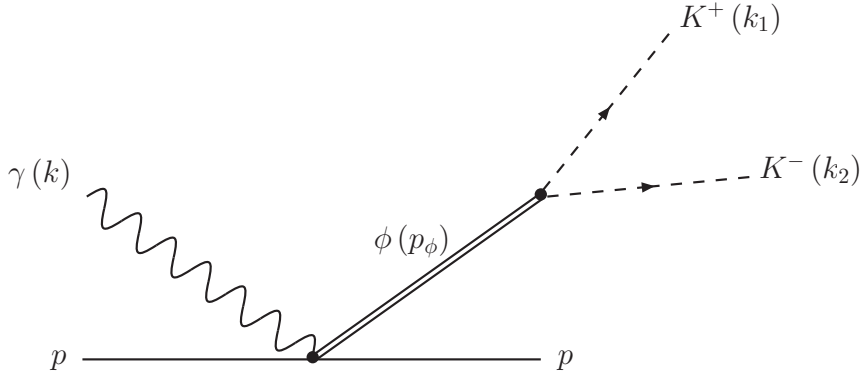}
\caption{Diagrammatic representation for the reaction $\gamma p\to\phi p\to K^{+}K^{-}p$.}\label{phiprodpropa}
\end{figure}
Once  the $\phi$ is produced, it can decay into a $K^{+}K^{-}$ pair as shown in Fig. \ref{phiprodpropa}. Therefore, naming $k_{1}$ and $k_{2}$ the four momentum of the $K^{+}$ and $K^{-}$, respectively, coming from the $\phi$ decay, we can describe the process $\gamma N\to \phi N\to K^{+}K^{-}N$ by means of the amplitude

\begin{align}
t_{\gamma N\to K^{+}K^{-} N}=\tilde{a}_{N}\epsilon^{\mu}_{\gamma}{\epsilon_{\mu\,\phi}}\frac{1}{p^{2}_{\phi}-M^{2}_{\phi}+iM_{\phi}\Gamma_{\phi}}g\epsilon^{\nu}_{\phi}(k_{1}-k_{2})_{\nu}
\end{align}
with $\Gamma_{\phi}$ the width of the $\phi$ and $g$ the $\phi K^{+}K^{-}$ coupling. Summing over the polarizations of the $\phi$, $\lambda_{\phi}$, and using that
\begin{align}
\sum_{\lambda_{\phi}}\epsilon^{\mu}_{\phi}\epsilon^{\nu}_{\phi}=-g^{\mu\nu}+\frac{p^{\mu}_{\phi}p^{\nu}_{\phi}}{M^{2}_{\phi}}
\end{align}
we have
\begin{align}
t_{\gamma N\to K^{+}K^{-} N}=a_{N}\frac{1}{(k_{1}+k_{2})^{2}-M^{2}_{\phi}+iM_{\phi}\Gamma_{\phi}}\epsilon^{\mu}_{\gamma}(k_{1}-k_{2})_{\mu}\label{ampli2}
\end{align}
where $a_{N}=\tilde{a}_{N}g$ and we have used that $p_{\phi}=k_{1}+k_{2}$ and taken an average mass for the kaons, i.e., $k^{2}_{1}=k^{2}_{2}=m^{2}_{K}$.

In the Coulomb gauge,  Eq. (\ref{ampli2}) reads as
\begin{align}
t_{\gamma N\to K^{+}K^{-} N}\equiv\vec{t}_{\gamma N\to K^{+}K^{-} N}\vec{\epsilon}_{\gamma}
\end{align}
with
\begin{align}
\vec{t}_{\gamma N\to K^{+}K^{-} N}\equiv -a_{N}\frac{1}{(k_{1}+k_{2})^{2}-M^{2}_{\phi}+iM_{\phi}\Gamma_{\phi}}(\vec{k}_{1}-
\vec{k}_{2})
\end{align}
and this is the expression which we use in order to describe the process $\gamma N\to \phi N\to K^{+}K^{-}N$. Note that the values of $\vec{k}_{1}$ and $\vec{k}_{2}$ depend on the diagram under evaluation.

For the one body process $k_{1}$, $k_{2}$ correspond to external variables which are known in the integration over phase space. The two body contributions are relatively smaller, even when the $\phi$ cut is done, as will be shown in section \ref{resca}. There one of the momentum is an external momentum while the other one corresponds to $\vec{q}_{ex}$, an integration variable, which we approximate in the matrix element by its value for $\vec{p}_{1}=\vec{p}_{2}=0$ in the laboratory frame, where the deuteron wave function in momentum space has its largest value. In this form, the amplitude $t_{\gamma N\to K^{+}K^{-}N}$ can be taken out of the integrals in Eqs. (\ref{Sc_d})-(\ref{Sf}).

The $t$-matrix for the $\bar{K}N$ and $K N$ interactions are calculated by solving the Bethe-Salpeter equation in a coupled channels approach, taking as potential the lowest order chiral amplitude as done in \cite{ramos,bennhold,felipe}. For the $\bar{K}N$ system and coupled channels, the scattering amplitude calculated with this approach reproduces the total cross section of $K^{-}p$ to all channels, and gives rise to  two $\Lambda(1405)$ states as dynamical generated resonances in the meson-baryon interaction \cite{meissner,meissner2}, which can be seen in the invariant mass distribution of $\pi\Sigma$ in different reactions \cite{meissner,jido}. The $t$ matrices for the processes $\bar{K}N\to \bar{K}N$ and $KN\to KN$ depend on the invariant mass of the interacting particles, which are given by the following expressions according to the diagrams of Fig. \ref{model} considered

\begin{align}
s^{(c)}&=(p_{n}+q^{\,\prime})^{2}=M^{2}_{n}+m^{2}_{K}+2p^{0}_{n}{{q}^{\,\prime}}^{0}-2\vec{p}_{n}\vec{q}^{\,\prime},\nonumber\\
s^{(d)}&=(p_{p}+q^{\,\prime})^{2}=M^{2}_{p}+m^{2}_{K}+2p^{0}_{p}{{q}^{\,\prime}}^{0}-2\vec{p}_{p}\vec{q}^{\,\prime},\nonumber\\
s^{(e)}&=(p_{n}+q)^{2}=M^{2}_{n}+m^{2}_{K}+2p^{0}_{n}{q}^{0}-2\vec{p}_{n}\vec{q},\\
s^{(f)}&=(p_{p}+q)^{2}=M^{2}_{p}+m^{2}_{K}+2p^{0}_{p}{q}^{0}-2\vec{p}_{p}\vec{q}.\nonumber
\end{align}

Since these invariant masses are defined in terms of external variables, the $\bar{K}N$ and $KN$ amplitudes can be taken out of the integrals in Eqs. (\ref{Sc_d})-(\ref{Sf}).

In these way we can define the following function
\begin{align}
\phi(q^{0}_{ex},\vec{Q})=\int \frac{d^{3}q_{ex}}{(2\pi)^{3}}\tilde{\varphi}(\vec{Q}-\vec{q}_{ex})\frac{1}{(q^{0}_{ex})^{2}-(\vec{q}_{ex})^{2}-m^{2}_{K}+i\epsilon}
\end{align}
which depends upon $q^{0}_{ex}$ and $\vec{Q}^{2}$, such that from Eqs. (\ref{Sc_d})-(\ref{Sf}) and Eq. (\ref{ST}) we can write the $T$ matrix from the different diagrams in Fig. \ref{model} as
\begin{align}
T^{(a)}&=\tilde{\varphi}\Bigg(\vec{p}_{n}-\frac{\vec{p}_{d}}{2}\Bigg)
 \vec{t}^{\,\,(a)}_{\gamma p\to K^{+}K^{-}p}\vec{\epsilon}_{\gamma}\equiv \overrightarrow{T}^{(a)}\vec{\epsilon}_{\gamma},
\nonumber\\
T^{(b)}&=\tilde{\varphi}\Bigg(\frac{\vec{p}_{d}}{2}-\vec{p}_{p}\Bigg)
\vec{t}^{\,\,(b)}_{\gamma n\to K^{+}K^{-}n}\vec{\epsilon}_{\gamma} \equiv \overrightarrow{T}^{(b)}\vec{\epsilon}_{\gamma},
\nonumber\\
T^{(c)}&=\phi\Bigg(k^{0}+p^{0}_{1}-q^{0}-p^{0}_{p},\vec{q}^{\,\prime}+\vec{p}_{n}-\frac{\vec{p}_{d}}{2}\Bigg)\vec{t}^{\,\,(c)}_{\gamma p\to K^{+}K^{-}p} t^{(c)}_{K^{-}n\to K^{-}n}\vec{\epsilon}_{\gamma} \equiv \overrightarrow{T}^{(c)}\vec{\epsilon}_{\gamma}\label{Ti},\\
T^{(d)}&=\phi\Bigg(k^{0}+p^{0}_{2}-q^{0}-p^{0}_{n},\vec{q}^{\,\prime} +\vec{p}_{p}-\frac{\vec{p}_{d}}{2}\Bigg)\vec{t}^{\,\,(d)}_{\gamma n\to K^{+}K^{-}n} t^{(d)}_{K^{-}p\to K^{-}p}\vec{\epsilon}_{\gamma} \equiv \overrightarrow{T}^{(d)}\vec{\epsilon}_{\gamma},\nonumber\\
T^{(e)}&=\phi\Bigg(k^{0}+p^{0}_{1}-{q^{\,\prime}}^{0}-p^{0}_{p},\vec{q}+\vec{p}_{n}-\frac{\vec{p}_{d}}{2}\Bigg)\vec{t}^{\,\,(e)}_{\gamma p\to K^{+}K^{-}p} t^{(e)}_{K^{+}n\to K^{+}n}\vec{\epsilon}_{\gamma} \equiv \overrightarrow{T}^{(e)}\vec{\epsilon}_{\gamma},\nonumber\\
T^{(f)}&=\phi\Bigg(k^{0}+p^{0}_{2}-{q^{\,\prime}}^{0}-p^{0}_{n},\vec{q}+\vec{p}_{p}-\frac{\vec{p}_{d}}{2}\Bigg)\vec{t}^{\,\,(f)}_{\gamma n\to K^{+}K^{-}n} t^{(f)}_{K^{+}p\to K^{+}p}\vec{\epsilon}_{\gamma} \equiv \overrightarrow{T}^{(f)}\vec{\epsilon}_{\gamma}.\nonumber
\end{align}
The total $T$ matrix will be given by the sum of  $T^{(i)}$, $i=a,b,\dots,f$. We have then from the different terms in Eqs. (\ref{Ti})
\begin{align}
T\equiv\overrightarrow{T}\vec{\epsilon}_{\gamma}
\end{align}
where 
\begin{align}
\overrightarrow{T}=\sum_{i=a}^{f}\overrightarrow{T}^{(i)}.
\end{align}

\subsection{Evaluation of the cross section}
In order to calculate the total cross section for the process $\gamma p\to K^{+}K^{-}pn$ we are going to work in the center of mass system, where we denote the variables with a tilde. In this frame, the cross section will be given by

\begin{align}
\sigma&=\frac{M_{d}}{s-M_{d}^{2}}\int\frac{d^{3}\tilde{p}_{n}}{(2\pi)^{3}}\int\frac{d^{3}\tilde{q}}{(2\pi)^{3}}\int\frac{d^{3}\tilde{q}^{\,\prime}}{(2\pi)^{3}}\int\frac{d^{3}\tilde{p}_{p}}{(2\pi)^{3}}\frac{1}{2\tilde{\omega}(\vec{\tilde{q}}\,\,)}\frac{1}{2\tilde{\omega}^{\,\prime}(\vec{\tilde{q}}^{\,\prime})}\frac{M_{p}}{\tilde{E}_{p}(\vec{\tilde{p}}_{p})}\frac{M_{n}}{\tilde{E}_{n}(\vec{\tilde{p}}_{n})}\nonumber\\
&\quad\times (2\pi)^{4}\delta^{4}(\tilde{k}+\tilde{p}_{d}-\tilde{p}_{p}-\tilde{p}_{n}-\tilde{q}-\tilde{q}^{\,\prime})\overline{\sum_{\lambda}}
|T|^{2}
\end{align}
with $M_{d}$ the deuteron mass, $\tilde{\omega}(\vec{\tilde{q}}\,\,)=\sqrt{\vec{\tilde{q}}^{\,\,2}+m^{2}_{K}}$, $\tilde{\omega}^{\,\prime}(\vec{\tilde{q}}^{\,\prime})=\sqrt{(\vec{\tilde{q}}^{\,\prime})^{2}+m^{2}_{K}}$, $\tilde{E}_{p}(\vec{\tilde{p}}_{p})=\sqrt{(\vec{\tilde{p}}_{p})^{2}+M^{2}_{N}}$ and $\tilde{E}_{n}(\vec{\tilde{p}}_{p})=\sqrt{(\vec{\tilde{p}}_{n})^{2}+M^{2}_{N}}$. We are using an average mass for the proton and neutron, i.e., $M_{p}=M_{n}=M_{N}$ where $M_{N}=(M_{p}+M_{n})/2$. Since the momentum of the neutron is not measured at LEPS, we can use the delta of Dirac to remove the integration in this variable. In this way we have
\begin{align}
\sigma&=\frac{M_{d}}{s-M_{d}^{2}}\int\frac{d^{3}\tilde{q}}{(2\pi)^{3}}\int\frac{d^{3}\tilde{q}^{\,\prime}}{(2\pi)^{3}}\int\frac{d^{3}\tilde{p}_{p}}{(2\pi)^{3}}\frac{1}{2\tilde{\omega}(\vec{\tilde{q}}\,\,)}\frac{1}{2\tilde{\omega}^{\,\prime}(\vec{\tilde{q}}^{\,\prime})}\frac{M_{p}}{\tilde{E}_{p}(\vec{\tilde{p}}_{p})}\frac{M_{n}}{\tilde{E}_{n}(-\vec{\tilde{p}}_{p}-\vec{\tilde{q}}-\vec{\tilde{q}}^{\,\prime})}\nonumber\\
&\quad\times(2\pi)\delta(\sqrt{s}-\tilde{\omega}(\vec{\tilde{q}}\,\,)-\tilde{\omega}^{\,\prime}(\vec{\tilde{q}}^{\,\prime})-\tilde{E}_{p}(\vec{\tilde{p}}_{p})-\tilde{E}_{n}(-\vec{\tilde{p}}_{p}-\vec{\tilde{q}}-\vec{\tilde{q}}^{\,\prime}))\overline{\sum_{\lambda}}|T|^{2}\label{delta}
\end{align}
where we have used that in the center of mass frame $\vec{\tilde{k}}+\vec{\tilde{p}}_{d}=0$.  The $\delta$ function is used to integrate over the angle between $\vec{\tilde{p}}_{p}$ and $\vec{\tilde{q}}+\vec{\tilde{q}}^{\,\prime}$ which is then given by
\begin{align}
cos \tilde{\theta}_{0}\equiv\frac{1}{2|\vec{\tilde{p}}_{p}|\cdot |\vec{\tilde{q}}+\vec{\tilde{q}}^{\,\prime}|}\Bigg[
(\sqrt{s}-\tilde{\omega}(\vec{\tilde{q}}\,\,)-\tilde{\omega}^{\,\prime}(\vec{\tilde{q}}^{\,\prime})-\tilde{E}_{p}(\vec{\tilde{p}}_{p}))^{2}-M^{2}_{N}-(\vec{\tilde{p}}_{p})^{2}-(\vec{\tilde{q}}+\vec{\tilde{q}}^{\,\prime})^{2}\Bigg]\nonumber
\end{align}
and the remaining integrals give now the cross section, which can be written as
\begin{align}
\sigma&=\frac{M_{d}M_{n}M_{p}}{s-M_{d}^{2}}\frac{1}{4(2\pi)^{2}}\int \frac{d^{3}\tilde{q}}{(2\pi)^{3}\tilde{\omega}(\vec{\tilde{q}}\,\,)}\int \frac{d^{3}\tilde{q}^{\,\prime}}{(2\pi)^{3}\tilde{\omega}^{\,\prime}(\vec{\tilde{q}}^{\,\prime})}\int d\tilde{E}_{p}\int d\tilde{\phi}\frac{1}{| \vec{\tilde{q}}+\vec{\tilde{q}}^{\,\prime}|}\overline{\sum_{\lambda}}|T|^{2}\nonumber\\
&\quad\times\Theta(1-cos^{2}\tilde{\theta}_{0})\Theta(\sqrt{s}-\tilde{\omega}(\vec{\tilde{q}}\,\,)-\tilde{\omega}^{\,\prime}(\vec{\tilde{q}}^{\,\prime})-\tilde{E}_{p}(\vec{\tilde{p}}_{p}))\label{cross}
\end{align}
where the $\Theta$ function guaranties that $|cos\tilde{\theta}_{0}|\leqslant 1$ and $\sqrt{s}-\tilde{\omega}(\vec{\tilde{q}}\,\,)-\tilde{\omega}^{\,\prime}(\vec{\tilde{q}}^{\,\prime})-\tilde{E}_{p}(\vec{\tilde{p}}_{p})>0$, as it should be.  In the Coulomb gauge, the element
\begin{align}
\overline{\sum_{\lambda}}|T|^{2}
\end{align}
can be written using the sum over photon polarizations
\begin{align}
{\sum_{\lambda}} {\epsilon}^{\,\,i}_{\gamma}{\epsilon}^{\,\,j}_{\gamma}=\delta_{ij}-\frac{k_{i}k_{j}}{\vec{k}^{2}}\label{eps}
\end{align}
as
\begin{align}
\overline{\sum_{\lambda}}|T|^{2}=\frac{1}{2}\Bigg[\overrightarrow{T}\cdot\overrightarrow{T}-\frac{(\vec{k}\cdot\overrightarrow{T})^{2}}{\vec{k}^{2}}\Bigg].
\end{align}
Although Eq. (\ref{cross}) corresponds to the cross section calculated in the center of mass frame, $\overline{\displaystyle{\sum_{\lambda}}}|T|^{2}$ is an invariant and can be calculated in any colinear frame. For this reason, for convenience, it is simpler to obtain it in the frame where the deuteron is at rest, i.e., $\vec{p}_{d}=0$. For this purpose, we have to make a boost from the center of mass frame to a frame which is moving with a speed 
\begin{align}
\vec{v}=\frac{\vec{k}+\vec{p}_{d}}{k^{0}+M_{d}}=\frac{\vec{k}}{k^{0}+M_{d}},
\end{align}
with $\vec{k}$, $k^{0}$ variables of the photon in the laboratory frame, according to the following relation \cite{pedro} 
\begin{align}
\vec{p}_{i}=\Bigg[\Bigg(\frac{k^{0}+M_{d}}{\sqrt{s}}-1\Bigg)\frac{\vec{\tilde{p}}_{i}\cdot\vec{k}}{\vec{k}^{2}}+\frac{\tilde{p}^{0}_{i}}{\sqrt{s}}\Bigg]\vec{k}+\vec{\tilde{p}}_{i}
\end{align}
with $\vec{\tilde{p}}$ an arbitrary vector  in the center of mass frame
and $\vec{p}$ the expression of this vector in the frame where the deuteron is at rest.

The integrals in Eq. (\ref{cross}) are done using the Monte Carlo method, which is particularly suited to implement the experimental cuts.
\subsection{$\Lambda(1520)$ production}\label{1520}
A detailed model for $\Lambda(1520)$ production, which reproduces fairly well the data, is available in \cite{sibirtsev}. In this work the Drell mechanism with $K^+$ and $K^-$ exchange is used to study the $\gamma p \to K^+ K^- p$  reaction, and the production of the $\Lambda(1520)$ is taken into account by a new term where the resonance is excited by means of  $K^*$ exchange.  Another phenomenological analysis is done in \cite{roca} where the $\Lambda(1520)$ resonance is assumed to be dynamically generated in coupled channels.  Here we follow a more empirical approach, since we are interested only about the shape of the $K^-p$ distribution and the strength of the $\Lambda(1520)$ production is fitted to the data. We assume that the $\Lambda(1520)$ does not interfere with the $\phi$ production mechanism, and we consider a minimal structure for $\gamma p\to K^{+}\Lambda(1520)\to K^{+}K^{-}p$ compatible with the quantum numbers of the $\Lambda(1520)$ ($J^{P}=3/2^{-}$) and the D-wave character \cite{roca}
\begin{align}
t_{\Lambda}=b_{\Lambda}(\vec{\sigma}\times\vec{q})\vec{\epsilon}\,\mathcal{D}_{\Lambda}\label{lamb}
\end{align}
with 
\begin{align}
\mathcal{D}_{\Lambda}=\frac{1}{M_{K^{-}p}-M_{\Lambda(1520)}+i\frac{\Gamma_{\Lambda(1520)}}{2}}
\end{align}
the $\Lambda(1520)$ propagator, $M_{\Lambda(1520)}$ and $\Gamma_{\Lambda(1520)}$ the mass and the width, respectively, of the $\Lambda(1520)$ resonance, $b_{\Lambda}$ a parameter, $\vec{\sigma}$ the Pauli matrices,  and $M_{K^{-}p}$ the $K^{-}p$ invariant mass. In this way
\begin{align}
|t_{\Lambda}|^{2}=|b_{\Lambda}|^{2}|\mathcal{D}_{\Lambda}|^{2}\epsilon_{ijk}\sigma_{j}q_{k}\epsilon_{i}\epsilon_{lmn}\sigma_{m}q_{n}\epsilon_{l}\label{tl}
\end{align}
Using Eq. (\ref{eps}), in the Coulomb gauge we find
\begin{align}
\overline{\sum_{\lambda_{\gamma},s}}|t_{\Lambda}|^{2}=\frac{1}{2}|b_{\Lambda}|^{2}|\mathcal{D}_{\Lambda}|^{2}\Bigg[\vec{q}^{\,\,2}+\frac{(\vec{k}\cdot\vec{q})^{2}}{\vec{k}^{2}}\Bigg].
\end{align}
We can add a background to this amplitude of the form $\frac{1}{2}|c_{\Lambda}|^{2}$, with $c_{\Lambda}$ being a parameter, and, then, substitute  in Eq. (\ref{cross})
\begin{align}
\overline{\sum_{\lambda}}|T|^{2}\to\overline{\sum_{\lambda}}|T|^{2}+\Bigg(\frac{1}{2}|b_{\Lambda}|^{2}|\mathcal{D}_{\Lambda}|^{2}\Bigg[\vec{q}^{\,\,2}+\frac{(\vec{k}\cdot\vec{q})^{2}}{\vec{k}^{2}}\Bigg]+\frac{1}{2}|c_{\Lambda}|^{2}\Bigg)\Bigg|\tilde{\varphi}\Bigg(\vec{p}_{n}-\frac{\vec{p}_{d}}{2}\Bigg)\Bigg|^{2}.
\end{align}
Since in the energy region considered the term
\begin{align}
\vec{q}^{\,\,2}+\frac{(\vec{k}\cdot\vec{q})^{2}}{\vec{k}^{2}}
\end{align}
does not change too much, we can take it as a constant and absorb it in the parameter $b_{\Lambda}$. This mean that in our case
\begin{align}
\overline{\sum_{\lambda}}|T|^{2}\to\overline{\sum_{\lambda}}|T|^{2}+\frac{1}{2}\Bigg(|b_{\Lambda}|^{2}|\mathcal{D}_{\Lambda}|^{2}+|c_{\Lambda}|^{2}\Bigg)\Bigg|\tilde{\varphi}\Bigg(\vec{p}_{n}-\frac{\vec{p}_{d}}{2}\Bigg)\Bigg|^{2}\label{eqtdos}.
\end{align}
We shall see that the results at LEPS can be reproduced with the $\phi$ and $\Lambda(1520)$ production, and there is no need for an extra background, represented by the $c_{\Lambda}$ term in the former equation.

The $\Lambda(1520)$ is dynamically generated in the $\bar{K}N$ system and coupled channels \cite{roca}. One can benefit from the study done in \cite{roca} to include the $\Lambda(1520)$ production in the $t$ matrix for the process $K^{-}p\to K^{-}p$ which appears in Fig. \ref{model}d. Following \cite{roca}
\begin{align}
t^{\Lambda(1520)}_{\bar{K}N\to\bar{K}N}&=t^{I=0}_{\bar{K}N\to\bar{K}N}\sum_{M}\mathcal{C}\Bigg(\frac{1}{2},2,\frac{3}{2}; m, M-m\Bigg)Y_{2, m-M}(\hat{p})\nonumber\\
&\quad\times\mathcal{C}\Bigg(\frac{1}{2},2,\frac{3}{2};m^{\prime},M-m^{\prime}\Bigg)Y_{2,m^{\prime}-M}(\hat{p}^{\prime})(-1)^{m^{\prime}-m}4\pi
\end{align}
with $\vec{p}$, $\vec{p}^{\,\prime}$ the initial and final momentum of the $K^{-}$, $t^{I=0}_{\bar{K}N\to\bar{K}N}$ the $t$ matrix for the $\bar{K}N$ system in isospin zero obtained solving the Bethe-Salpeter equations in a unitary chiral approach \cite{ramos,bennhold}, $\mathcal{C}$ Clebsch-Gordan coefficients and $Y$ the spherical harmonics. We take $\hat{p}$ in the $z$ direction, and then, the only not zero spherical harmonic is $Y_{2,0}(\hat{p})$.
In LEPS they measure angles in the forward direction. Then, taking $\hat{p}$ and $\hat{p}^{\prime}$ in the forward direction
\begin{align}
Y_{2,0}(\hat{p})=Y_{2,0}(\hat{p}^{\prime})=\sqrt{\frac{5}{4\pi}}.
\end{align}
Let us consider $m=m^{\prime}=1/2$.  Since $\mathcal{C}\Bigg(\frac{1}{2},2,\frac{3}{2};\frac{1}{2},0,\frac{1}{2}\Bigg)=\sqrt{\frac{2}{5}}$, we have that 
\begin{align}
t^{\Lambda(1520)}_{\bar{K}N\to\bar{K}N}=2t^{I=0}_{\bar{K}N\to\bar{K}N}.
\end{align}
Using that
\begin{align}
|\bar{K}N, I=0\rangle=\frac{1}{\sqrt{2}}|K^{-}p+\bar{K}^{0}n\rangle
\end{align}
we can determine the element $t^{\Lambda(1520)}_{K^{-}p\to K^{-}p}$ as
\begin{align}
t^{\Lambda(1520)}_{K^{-}p\to K^{-}p}=\frac{t^{I=0}_{\bar{K}N\to\bar{K}N}}{2}=t^{I=0}_{\bar{K}N\to\bar{K}N}
\end{align}
and close to the resonance, $t^{I=0}_{\bar{K}N\to\bar{K}N}$ can be written like
\begin{align}
t^{I=0}_{\bar{K}N\to\bar{K}N}=\frac{g^{2}_{\bar{K}N}}{M_{K^{-}p}-M_{\Lambda(1520)}+i\frac{\Gamma_{\Lambda(1520}}{2}}
\end{align}
where $g_{\bar{K}N}$ is the coupling of the $\Lambda(1520)$ to the $\bar{K}N$ channel, which according to \cite{roca} is -0.54.
This term must be added to the s-wave $K^{-}p\to K^{-}p$ amplitude which appears in Eqs. (\ref{Sd}). We mention already here that including this term to the rescattering part induces corrections which provide a small contribution, thus, has negligible effects in the final results.
\subsection{Set up at LEPS}\label{MMSA}
The LEPS detector is a forward magnetic spectrometer, therefore, it can only detect particles in the forward direction. The
angular coverage of the spectrometer is approximately $\pm$ 20 and $\pm$10 degrees in the horizontal
and the vertical directions, respectively. In our simulation we impose that the angle of the kaons in the final state with respect the incident photon is not bigger than 20 degrees. 

The nucleons are not detected at LEPS, therefore, some prescription is required in order to estimate the momentum of the $p$ and $n$ in the reaction  $\gamma d \to K^{+}K^{-}np$ and determine the invariant mass of $K^{-}p$ or $K^+ n$. This is done
using the minimum momentum spectator approximation (MMSA). The aim is to guarantee that a chosen nucleon is the closest possible to be a spectator in the reaction. The chosen spectator nucleon is the proton if one evaluates the $K^{+}n$ invariant mass, or the neutron when the $K^{-}p$ invariant mass is evaluated. For this purpose one defines the magnitude 
\begin{align}
p_{pn}=p_{miss}=p_{\gamma} +p_d - p_{K^+} -p_{K^-}
\end{align}
which corresponds to the four momentum of the outgoing $pn$ pair. In the $pn$ center of mass frame, the momentum of a nucleon is given by  
\begin{align}
|\vec{p}_{CM}| = \frac{\lambda^{1/2}(M^{2}_{pn},M^{2}_{p},M^{2}_{n})}{2M_{pn}}
\end{align}
Boosting back this momentum to the laboratory frame, we will have a minimum modulus for the momentum of the spectator nucleon when the momentum $\vec{p}_{CM}$ for this nucleon goes in the direction opposite to $\vec{p}_{miss}$. Thus, the minimum momentum, $p_{min}$, is given by 
 \begin{align}
 p_{min} = -|\vec{p}_{CM}| \cdot \frac{E_{miss}}{M_{pn}} + E_{CM} \cdot\frac{|\vec{p}_{miss}|}{M_{pn}}
\end{align}
where $E_{CM}=\sqrt{|\vec{p}_{CM}|^{2}+M^{2}_{N}}$ is the energy of the nucleon in the CM frame. In this case, the momentum of the other nucleon will be in the direction of the missing momentum  with a magnitude 
\begin{align}
p_{res} = |\vec{p}_{miss}| -p_{min}	
\end{align}
In \cite{nakatwo} the $M_{K^+ n}$ invariant mass for the reaction $\gamma d\to K^{+}K^{-}np$ is evaluated assuming the proton to have a momentum $p_{min}$. Consequently, in this prescription, the momentum of the neutron in the final state will be
\begin{align}
\vec{p}_{n}=p_{res}\cdot\frac{\vec{p}_{miss}}{|\vec{p}_{miss}|}
\end{align}
which is used to calculate the $M_{K^{+}n}$ invariant mass for the reaction $\gamma d\to K^{+}K^{-}np$ in \cite{nakatwo}.  At LEPS only events which satisfy $|p_{min}|<$ 100 MeV are considered. This condition is also implemented in our simulation of the process.

In order to remove the contribution from the $\phi$ production at LEPS one considers events which satisfy that the invariant mass of the $K^{+}K^{-}$ pair is bigger than 1030 MeV and bigger than the value obtained from the following expression
\begin{align}
1020\, \textrm{MeV} +0.09\times (E^{eff}_{\gamma}(\textrm{MeV})-2000\, \textrm{MeV})
\end{align}
where $E^{eff}_{\gamma}$ is defined as the effective photon energy
\begin{align}
E^{eff}_{\gamma}=\frac{s_{K^{+}K^{-}n}-M^{2}_{n}}{2M_{n}}
\end{align}
with $s_{K^{+}K^{-}n}$ the square of the total center of mass energy for the $K^{+}K^{-}n$ system calculated using the MMSA approximation to determine the momentum of the neutron assuming the proton as spectator. In \cite{nakatwo} only events for which $2000$ MeV $<E^{eff}_{\gamma} $ $<2500$ MeV are considered, a condition which is also incorporated in our simulation. This is the cut implemented when the invariant $K^{+}n$ mass is reconstructed. When the $K^{-}p$ mass is reconstructed, the cut is the same but $E^{eff}_{\gamma}$ is now evaluated with the MMSA prescription assuming the neutron as spectator. These energies would correspond to the photon in a $\gamma n$ ($\gamma p$) reaction with the original $p$ ($n$) at rest.

\section{Results with $\phi$ production}\label{penphi}
We consider first the production of $\phi$ on proton and neutron followed by rescattering and we look at the distribution of $d\sigma/d M_{K^{+}n}$ which is obtained taking for the neutron the momentum provided by the MMSA prescription described in the former section.

We have made the integrals in Eq. (\ref{cross}) using the Monte Carlo integration method with different seeds and with a statistics of 20 million points in the variables of Eq. (\ref{cross}), such that we get about 66500 ``good" points in the integral over the phase space which pass all the tests of the cuts. In order to take into account that the photon energies in the LEPS experiment run from 2 GeV to 2.4 GeV, we have generated 20 random energies in that range. 

\begin{figure}[h!]
\centering
\includegraphics[width=0.7\textwidth]{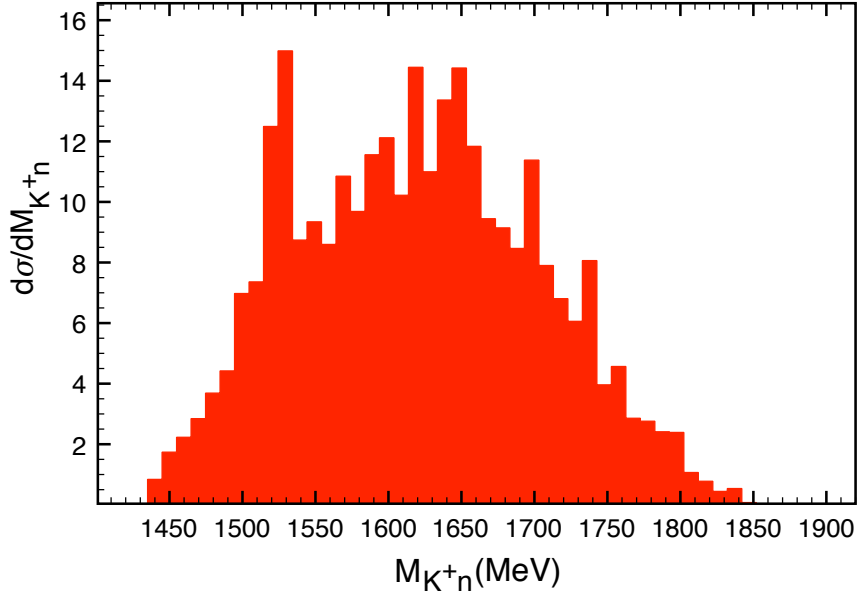}
\caption{(Color online) $M_{K^{+}n}$ invariant mass distribution with about 66500 ``good'' points.}\label{penta}
\end{figure}

In Fig. \ref{penta} we show the first distribution obtained with the cuts of \cite{nakatwo} and using the MMSA prescription for the nucleon momenta as done in \cite{nakatwo}. What we see in the figure is a pronounced peak around an invariant $K^{+}n$ mass of 1530 MeV, very similar to the peak observed in \cite{nakatwo}, which was identified as a possible signature of the $\Theta^{+}$ pentaquark. However, the peak has appeared here as a consequence of the cuts and not as a signal of a resonance which is not present in our theoretical model. 

One may rightly ask oneself whether the peak obtained in Fig. \ref{penta} is a necessary peak coming from the LEPS cuts or it is a statistical fluctuation. The answer comes from running with much more statistics, which we show below. However, before showing these results let us mention that statistically we get a clear peak around this energy in about $1/3$ of the runs with different seeds, but in all of them there is some trace of extra strength in this region.

\begin{figure}
\centering
\includegraphics[width=0.7\textwidth]{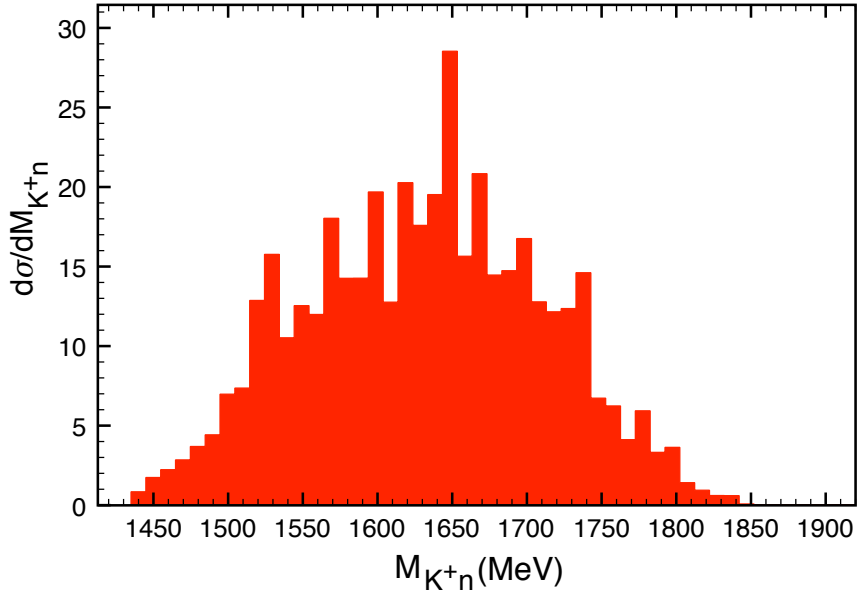}
\caption{$M_{K^{+}n}$ invariant mass distribution with about  71000 ``good'' points and with the cuts $p_{min}<$100 MeV and $M_{K\bar{K}}>$ 1030 MeV.}\label{pentades}
\end{figure}

Since one may argue that statistically it is more probable not to find a clear peak (about $2/3$ probability), we find opportune to make the following test. Let us imagine that we had devised that a proper cut to see a possible signature of the $\Theta^{+}$ would be to take 1030 MeV for the $\phi$ cut and 100 MeV for the $p_{min}$ test. The results with these cuts can be seen in Fig. \ref{pentades}.
A peak is still seen around $M_{K^{+}n}\sim$ 1530 MeV, but of the same size as the statistical  fluctuations that one observes in other parts of the spectrum. If in an eventual experiment one might decide that a cut like the one applied for Fig. \ref{penta} is more appropriate to analyze the experiment, one would then find the peak around 1530 MeV of Fig. \ref{penta} and could see in it the signature of a new resonance.

\begin{figure}
\centering
\includegraphics[width=0.7\textwidth]{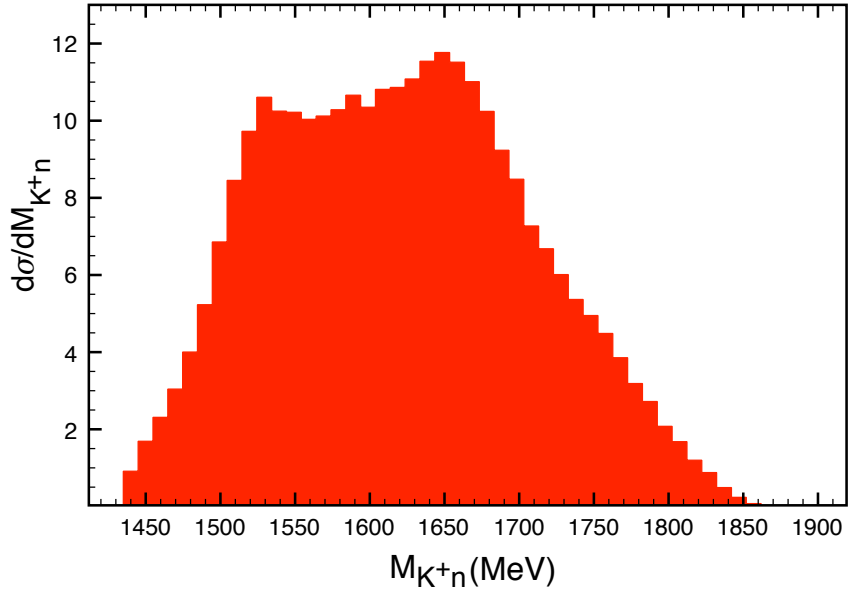}
\caption{(Color online) $M_{K^{+}n}$ invariant mass distribution calculated with $\sim$ 6420000 ``good'' points with the same cuts as those made in \cite{nakatwo} at LEPS.}\label{pentamore}
\end{figure}

After these exercises one still would like to see what happens with large statistics. We have made a run with 2000 millions points, out of which around 6420000 points are ``good" points passing the test of the LEPS cuts.  The results are shown in Fig. \ref{pentamore}. We see that the narrow peak has disappeared in the 1530 MeV region, but a softer bump still remains. A structure of this type in an eventual experimental spectrum could easily be interpreted as due to a resonance present in the reaction, but in our case it has been created by the cuts taken from the LEPS set up since we do not have any $K^{+}n$ resonance in the model.

\begin{figure}
\centering
\includegraphics[width=0.7\textwidth]{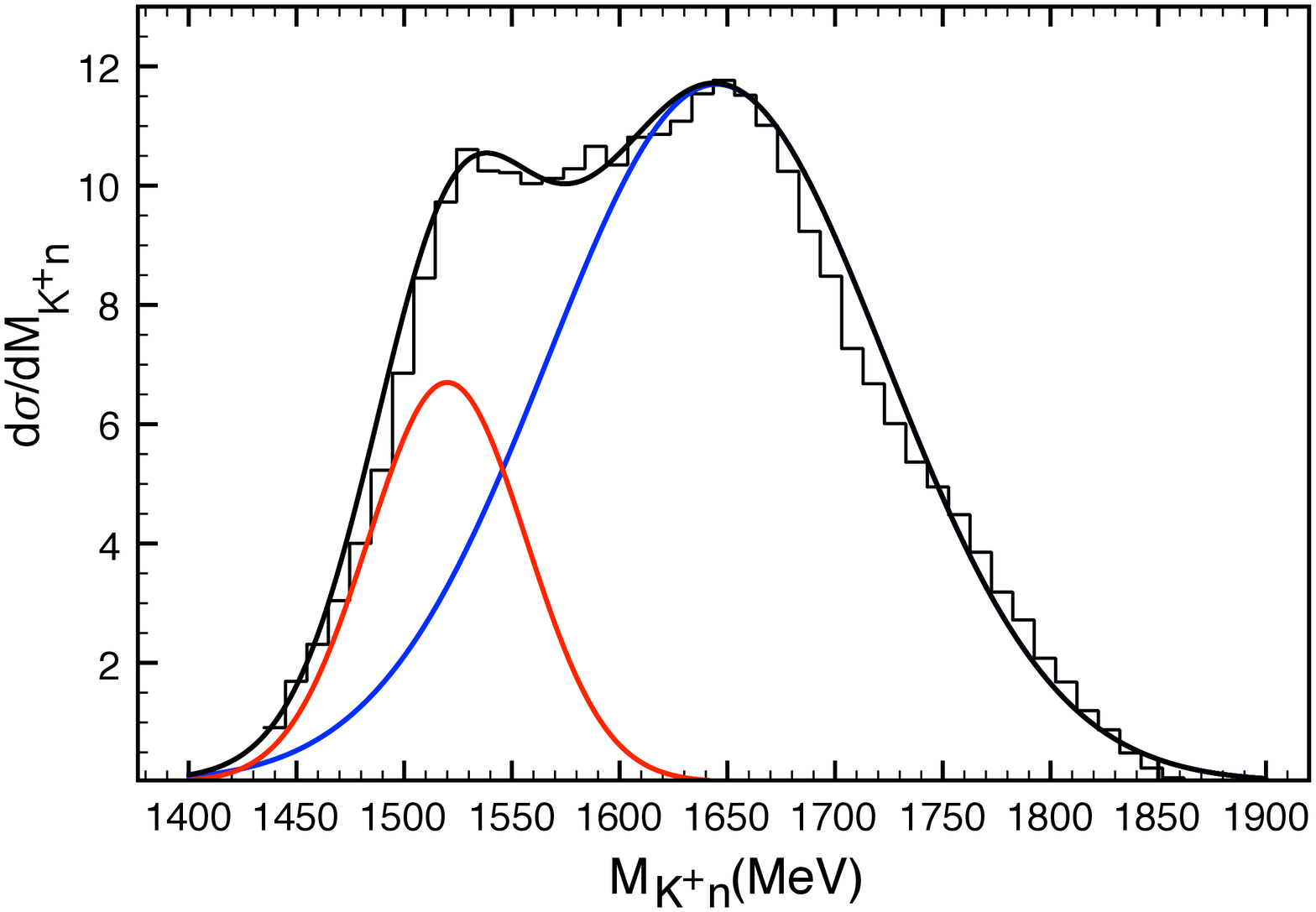}
\caption{(Color online) $M_{K^{+}n}$ invariant mass distribution shown in Fig. \ref{pentamore}, together with the result from the sum of two gaussian functions, one peaking at 1520 MeV with a width of 85 MeV and another one peaking at 1645 MeV with a width of 185 MeV.}\label{separa}
\end{figure}

 One can visualize the comment made above by separating the mass spectrum of Fig. \ref{pentamore} into a smooth background and a ``resonant" signal as shown in Fig.\ref{separa}, as is usually done in many experimental analyses. A structure with resonant shape around 1520 MeV and a width  of about 85 MeV has been generated as a consequence of the cuts. 
 
\begin{figure}
\centering
\includegraphics[width=0.7\textwidth]{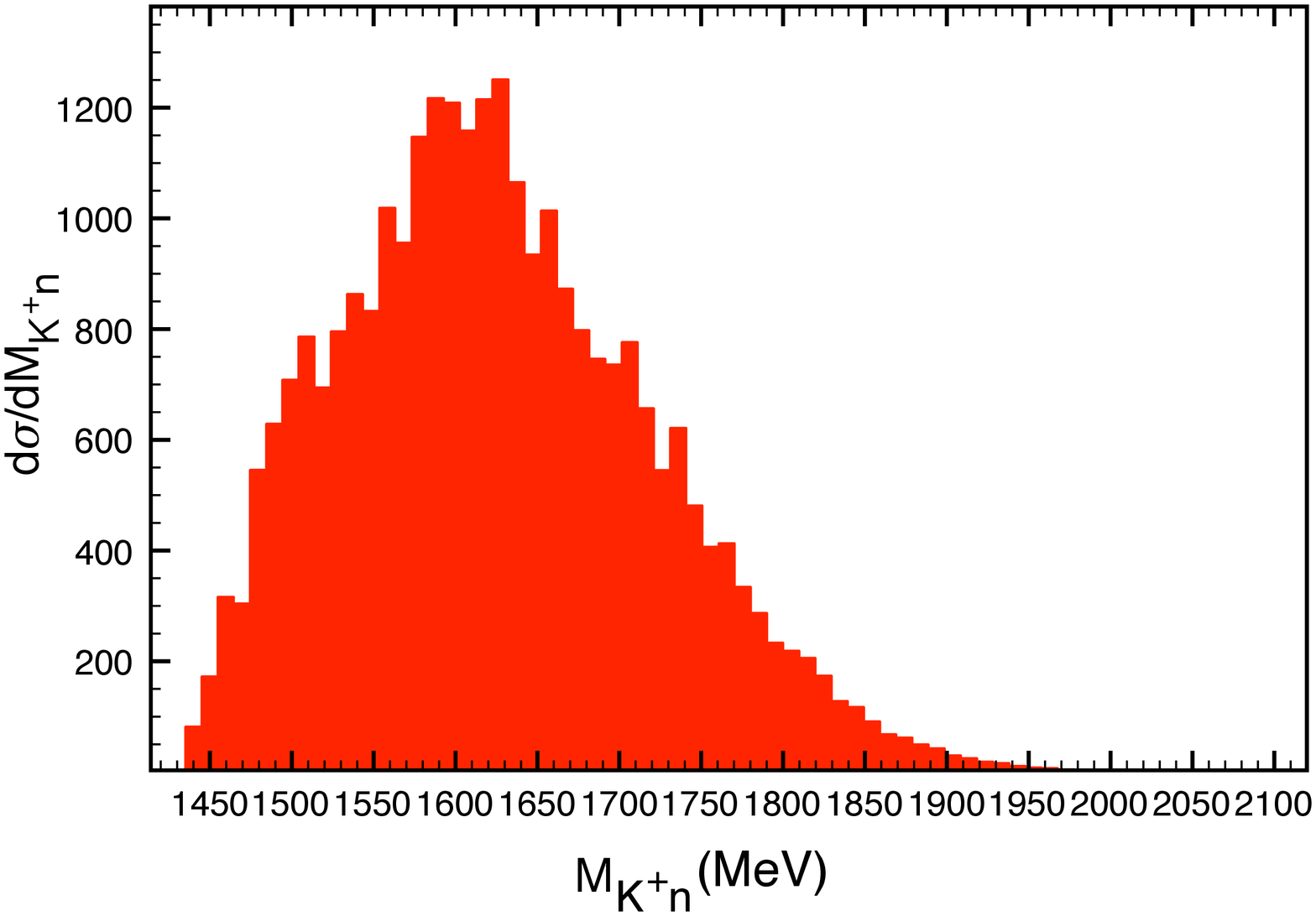}
\caption{(Color online) $M_{K^{+}n}$ invariant mass distribution obtained with $\sim$ 41280000 ``good'' points without any cut.}\label{kplusn_nocut}
\end{figure}

\begin{figure}
\centering
\includegraphics[width=0.7\textwidth]{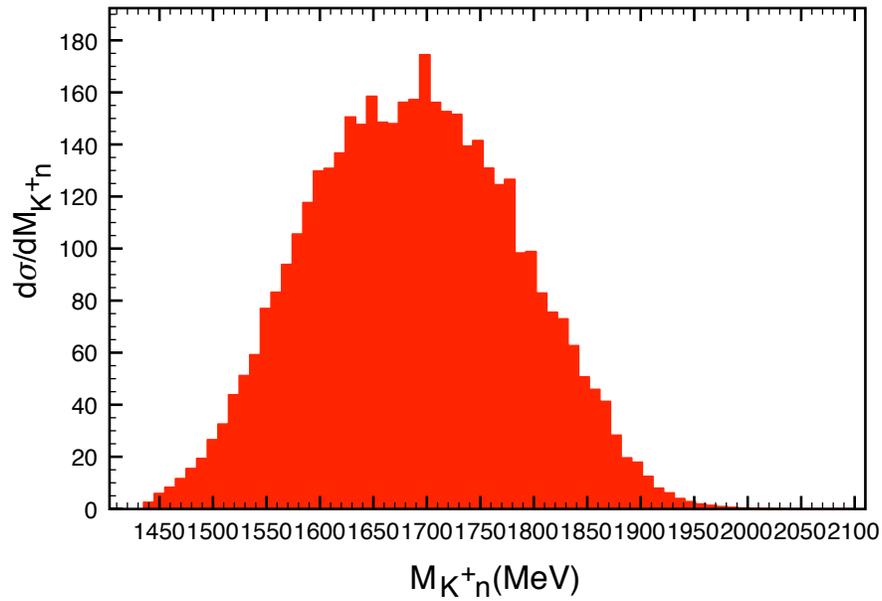}
\caption{(Color online) $M_{K^{+}n}$ invariant mass distribution calculated with $\sim$ 23900000 ``good'' points with only the angular cut.}\label{kplusn_anglecut}
\end{figure}

\begin{figure}
\centering
\includegraphics[width=0.7\textwidth]{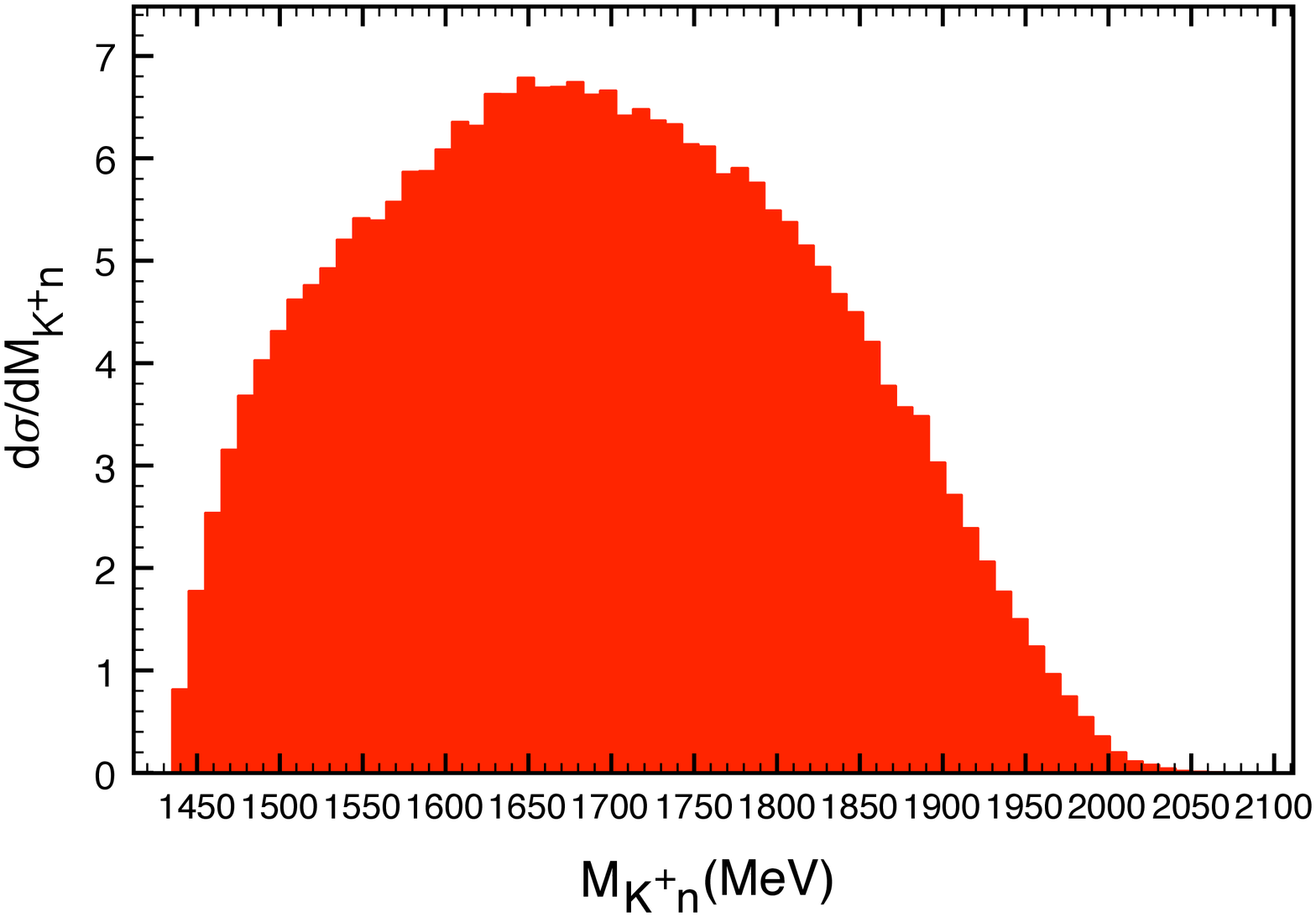}
\caption{(Color online) $M_{K^{+}n}$ invariant mass distribution obtained with $\sim$ 12700000 ``good'' events  with the $\phi$ cut $M_{K^{+}K^{-}}>$ 1050 MeV and the angular cut.}\label{kplusn_phicut_anglecut}
\end{figure}

In order to see the effects of the cuts we show in Fig. \ref{kplusn_nocut}, \ref{kplusn_anglecut}, \ref{kplusn_phicut_anglecut} the results for the distributions which we get by not putting any cut, only the angle cut, or the angle and a $\phi$ cut, i.e., $M_{K^{+}K^{-}}>$ 1050 MeV, respectively. In figures \ref{kplusn_nocut}, \ref{kplusn_anglecut} and \ref{kplusn_phicut_anglecut} the neutron momentum is the one provided in the Monte Carlo runs and is not evaluated according to the MMSA prescription as in the former figures. What we observe is that the ``resonant like'' structure of Figs. \ref{pentamore}, \ref{separa} does not show up in any of the runs where the real momenta have been used. Since the $\phi$ cut used in \cite{nakatwo} is dependent on $p_{min}$ and the MMSA prescription is used for the kinematics of the nucleons, we must conclude that the combination of these two facts is responsible for the ``resonant like'' structure seen in Figs. \ref{pentamore}, \ref{separa}.

\section{Introducing the $\Lambda(1520)$}\label{in1520}
Now we introduce in the amplitude $\gamma d\to K^{+}K^{-}np$ a term accounting for the production of the $\Lambda(1520)$ as explained in section \ref{1520}. The different structure in the momenta of the final states  for the $K^{+}\Lambda(1520)$ and $\phi p$ production suggest the incoherent sum of cross sections and we follow this option. 

\begin{figure}
\centering
\includegraphics[width=0.7\textwidth]{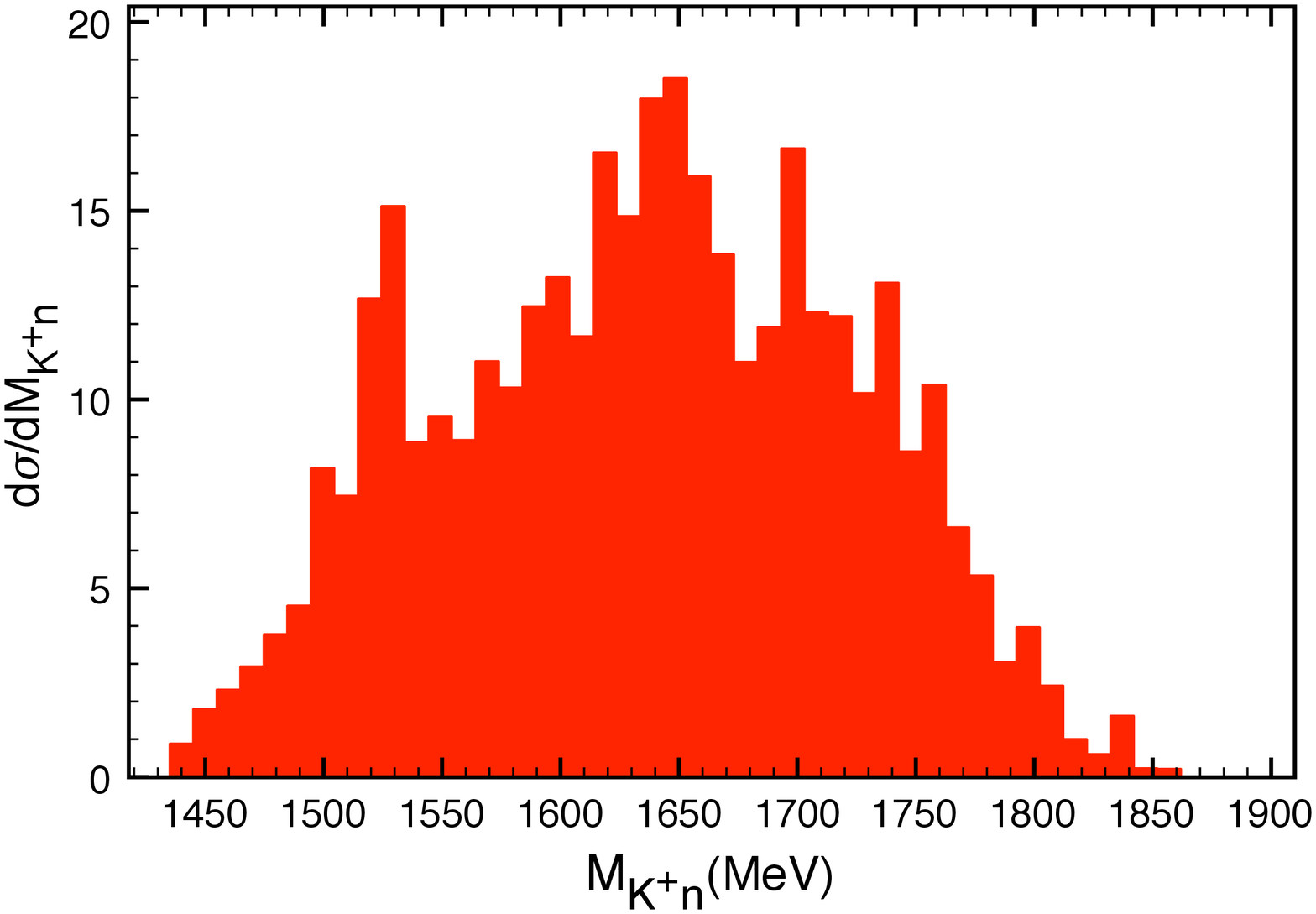}
\caption{(Color online) $M_{K^{+}n}$ invariant mass distribution obtained with $\sim$ 66500 ``good'' points, the same cuts as those made in \cite{nakatwo} at LEPS and with the $\Lambda(1520)$ term.}\label{pentaLamb}
\end{figure}

We find that adding the $\Lambda(1520)$ production in the amount demanded by the LEPS experiment does not spoil the mass distribution found in Fig. \ref{penta}, taking the same seed for the Monte Carlo integration. This is shown in Fig. \ref{pentaLamb}. 
However, the introduction of the $\Lambda(1520)$ production term serves us also another purpose. Indeed, we can make a test to see how the prescription used at LEPS to identify resonances works in this case. Since neither the final proton not the neutron are detected in the experiment, one must take some prescription to have the best guess for  the neutron momentum  in order to determine the $K^{+}n$ invariant mass of an event. The prescription taken at LEPS relies on the concept of minimum momentum as we explained in section \ref{MMSA}.
\begin{figure}
\centering
\includegraphics[width=0.7\textwidth]{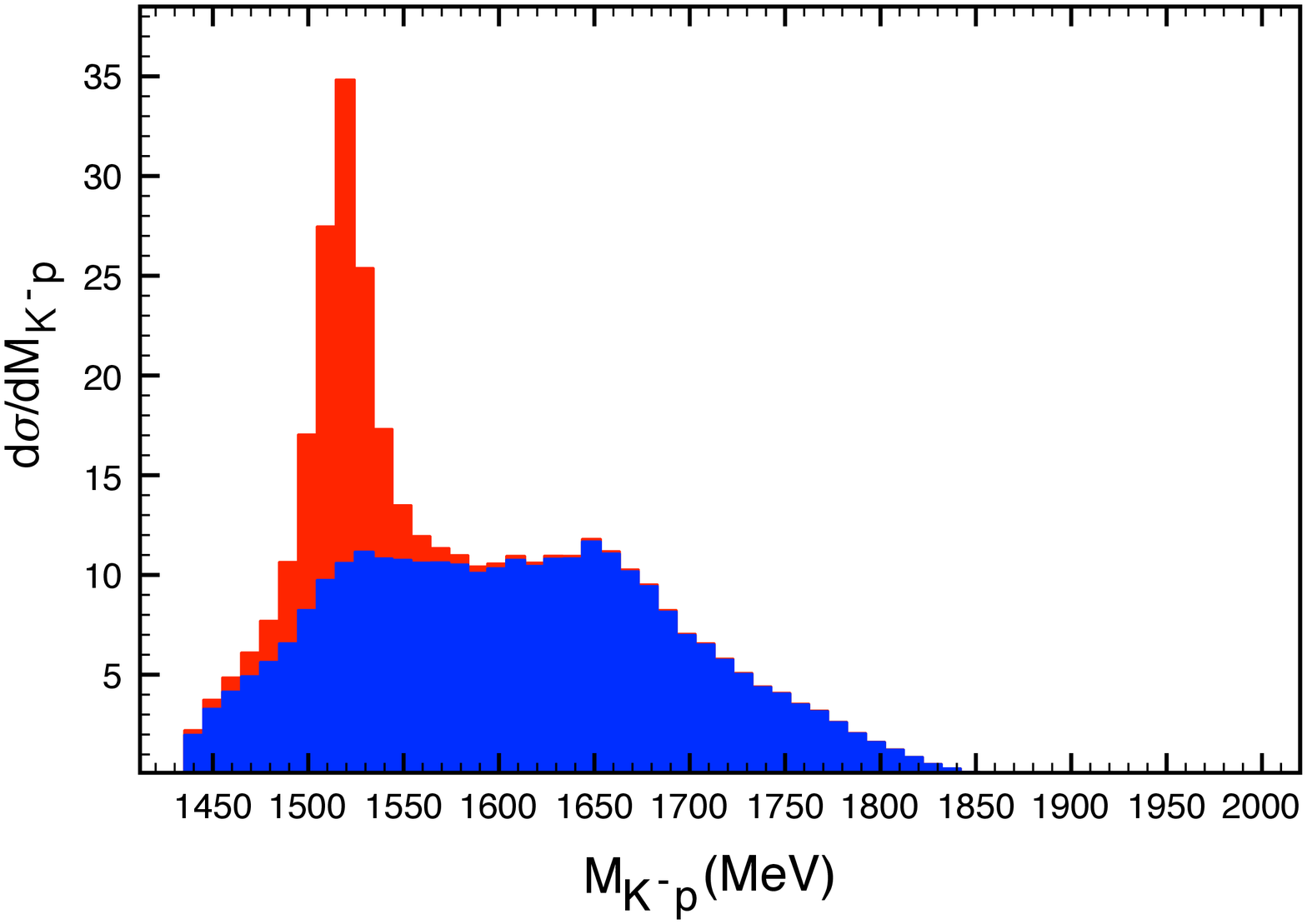}
\caption{(Color online) $M_{K^{-}p}$ invariant mass distributions with and without the $\Lambda(1520)$ term and making the same cuts that at LEPS. These distributions have been obtained with $\sim$ 6420000 ``good'' points.}\label{Lambda}
\end{figure}
In Fig. \ref{Lambda} we show the results for the $K^{-}p$ invariant mass considering the same cuts as in LEPS. We can see that, using the MMSA prescription to calculate the $K^{-}p$ invariant mass as described in section \ref{MMSA}, one gets a peak at the right mass coming from the $\Lambda(1520)$, and the spectrum resembles much the spectrum found at LEPS. In this case, if we remove the term of the $\Lambda(1520)$ production in the $\gamma p\to K^{+}K^{-}p$ amplitude and plot the $K^{-}p$ invariant mass distribution we do not find the signal of the resonance, as can be seen in Fig. \ref{Lambda} (lower area of the spectrum). It is interesting to note that, by reasons of symmetry, the spectrum obtained for the $K^- p$ invariant mass in the absence of any $\Lambda(1520)$ term is basically the same as the one obtained for the $K^+ n$ invariant mass in Fig. \ref{pentamore}, up to minor differences from the rescattering terms. It is interesting to point out here that, should one have a statistical fluctuation around 1530 MeV as one has in Fig. \ref{penta}, it would go unnoticed since it would be buried below the $\Lambda(1520)$ peak, which has become wider than the natural width as a consequence of the MMSA prescription.

\begin{figure}[h!]
\centering
\includegraphics[width=0.7\textwidth]{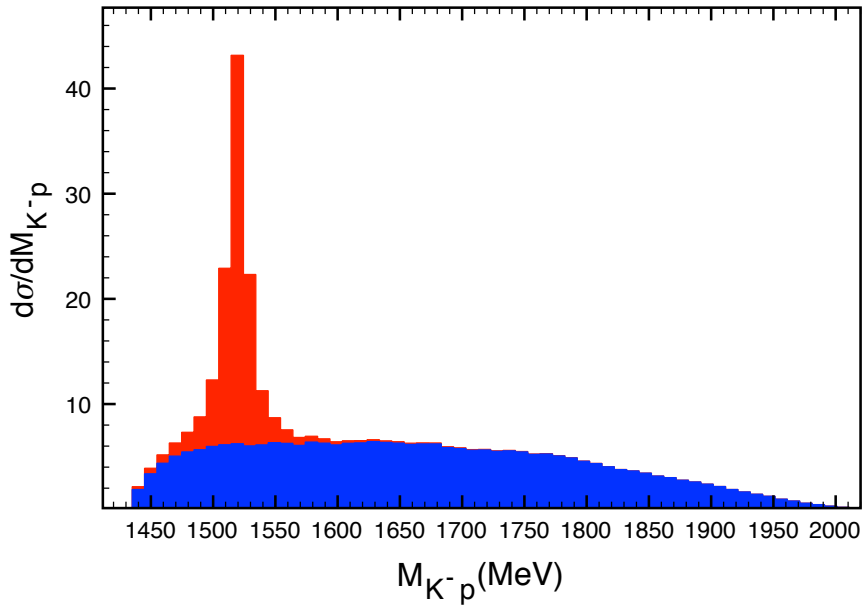}
\caption{(Color online) $M_{K^{-}p}$ invariant mass distributions with and without the $\Lambda(1520)$ term and making  the angle cut and with $M_{K^{+}K^{-}}>$ 1050 MeV using the real $K^{-}$ and $p$ momenta. These distributions have been calculated with $\sim$ 12700000 ``good'' points.}\label{nocuts}
\end{figure}

Fig. \ref{Lambda} shows, indeed, that the prescription to identify a resonance using the MMSA prescription works if the cross section for the production of this resonance is sufficiently large. However, let us see what would happen if we actually measured the proton such that we can clearly determine the $K^{-}p$ distribution without using the MMSA prescription. The results are seen in Fig. \ref{nocuts}. We can see that the peak for the $\Lambda(1520)$ is sharper than in Fig. \ref{Lambda}. We also notice that the non $\Lambda(1520)$ part of the spectrum (lower area in the figure) changes from one figure to the other. This is telling us that the MMSA prescription can indeed be used to identify a resonance, but it also distorts the shape of the spectrum and could have some repercussions in the $K^+ n$ spectrum, as we have already seen.  We have checked that the contribution to the spectrum at large $K^{-}p$ invariant masses in Fig. \ref{nocuts} comes from $\phi$ production on the neutron, where the proton is a spectator.

\begin{figure}[h!]
\centering
\includegraphics[width=0.7\textwidth]{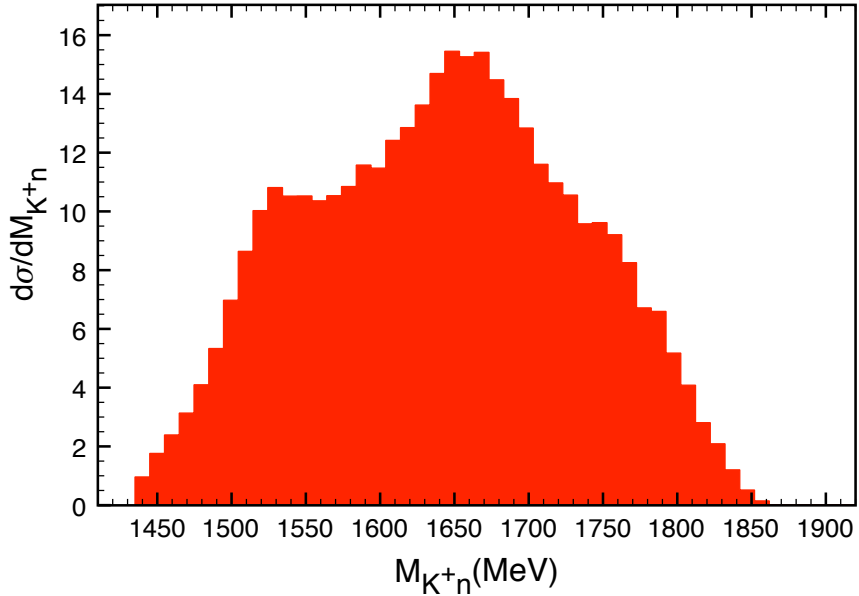}
\caption{(Color online) $M_{K^{+}n}$ distribution calculated with the same cuts as those done in \cite{nakatwo} and including the $\Lambda(1520)$ term for $\sim$ 6420000 ``good'' points.}\label{pentamore_Lambda}
\end{figure}

Before ending the section, we show in Fig. \ref{pentamore_Lambda} the equivalent of Fig. \ref{pentamore} using the MMSA prescription and the cuts done in \cite{nakatwo}, but with the full model, i.e., including $\phi$ and $\Lambda(1520)$ production. The results resemble those of Fig. \ref{pentamore} but the peak at higher invariant masses is a bit more pronounced. This should be compared to Fig. 12 of \cite{nakatwo}.

\section{Rescattering contribution}\label{resca}
\begin{figure}[h!]
\centering
\includegraphics[width=0.7\textwidth]{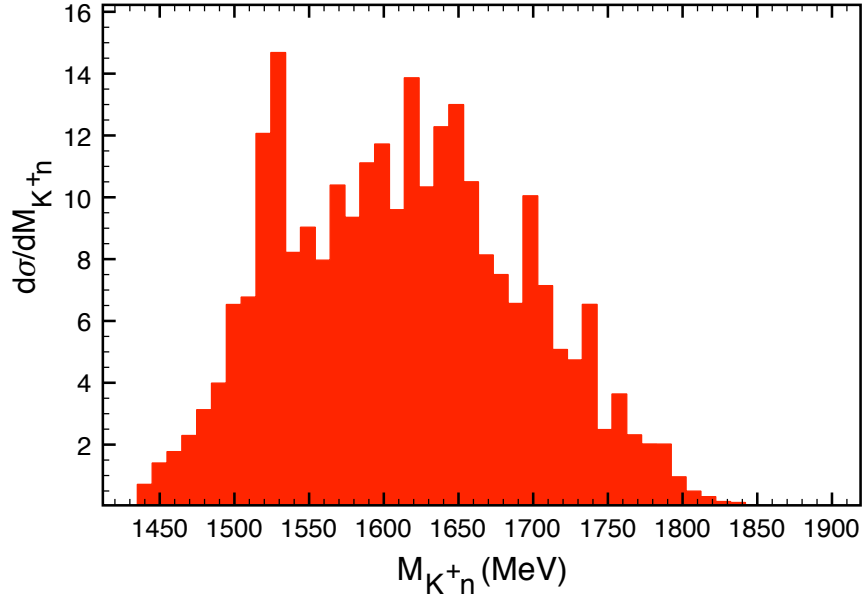}
\caption{(Color online) $M_{K^{+}n}$ invariant mass distribution with a statistic similar to the one in Fig. \ref{penta}, including only the single scattering contribution and with the same cuts as those done in \cite{nakatwo} at LEPS.}\label{single}
\end{figure}

\begin{figure}[h!]
\centering
\includegraphics[width=0.7\textwidth]{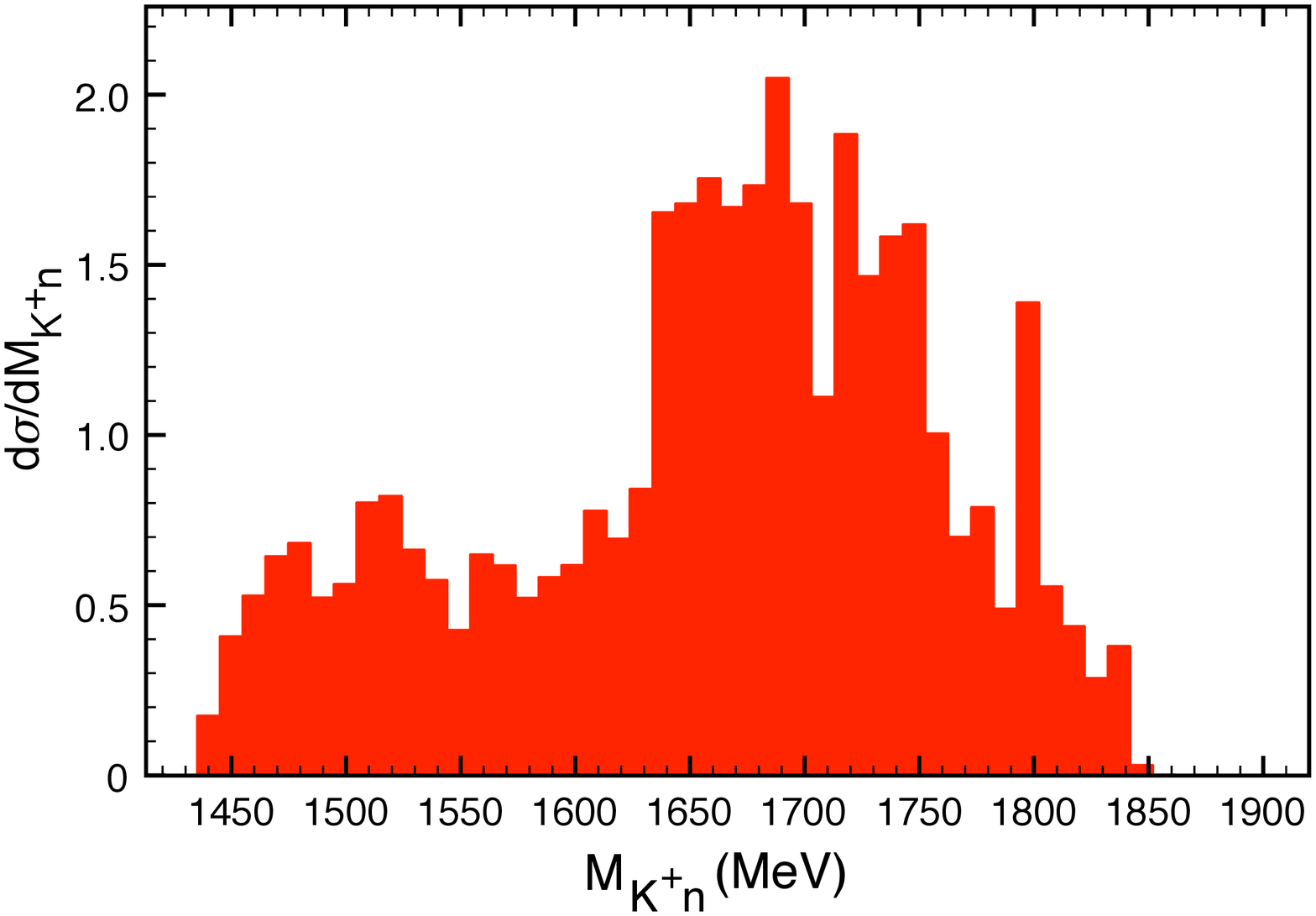}
\caption{(Color online) $M_{K^{+}n}$ invariant mass distribution with a statistic similar to the one in Fig. \ref{penta}, including only the rescattering contribution and with the same cuts as those done in \cite{nakatwo} at LEPS.}\label{double}
\end{figure}

In Fig. \ref{single} we show the $M_{K^{+}n}$ distribution calculated with the MMSA prescription and the same cuts as those made in \cite{nakatwo} at LEPS obtained considering only the contribution from the single scattering, 
which corresponds to the diagrams Fig. \ref{model}a, \ref{model}b, while in Fig. \ref{double} we show the result obtained taking into account only the rescattering diagrams of Fig. \ref{model}.

As one can see, the single scattering terms provide the dominant contribution, although the double scattering terms help
in increasing a bit the magnitude and their contribution also shows some peak around 1520 MeV.

\section{Statistical analysis and comparison with LEPS results}\label{sta1}
 In this section we address the question of statistics and error analysis. In the first place we consider  the error from the Monte Carlo simulation. This is given by the standard formula

 \begin{align}
\epsilon^{2}_{r}=\frac{1}{\displaystyle{\sum_{j=1}^{N_{E}}n_{j}}}\left\{\frac{\displaystyle{\Bigg(\sum_{j=1}^{N_{E}}n_{j}\Bigg)^{-1}\sum_{j=1}^{N_{E}}\sum_{i=1}^{n_{j}}V^{2}_{j}f^{2}}}{\Bigg [\Bigg(\displaystyle{\sum_{j=1}^{N_{E}}n_{j}\Bigg)^{-1}}\displaystyle{\sum_{j=1}^{N_{E}}\sum_{i=1}^{n_{j}}V_{j}f\Bigg]^{2}}}-1\right\} \label{formuguar}
\end{align}
where $\epsilon_{r}$ is the relative error for $d\sigma/d M_{inv}$ in a bin of the invariant mass, $f$ is the integrand in the integral for the cross section (Eq. (\ref{cross})) for the Monte Carlo points that fall inside a certain bin, $N_{E}$  the number of energies of the photon for which we calculate the distributions, $n_{j}$ the number of ``good'' events in one bin for each of these energies and $V_{j}$ the volume of the phase space for each photon energy.

Coming back to Fig. \ref{penta}, we obtain using Eq. (\ref{formuguar}) a relative error of $\sim$ 20 \% in the bins of invariant mass around 1530 MeV. This is an error that justifies the fluctuation seen in the figure.

 The equivalence of $N$ integration points and $\tilde{N}$ measured events in a experiment from the statistical point of view is not straightforward. For this reason, we approach now a different procedure which is suited to generate events like in the experiment, which we call simulation run. We use the Von Neumann rejection method to generate events. This procedure works like a Quantum Mechanical experiment in which events are produced or not according to their probability. The Von Neumann rejection method does that: it generates or not an event according to its probability measured with respect to the maximum probability, where the probability is given by the integrand of Eq. (\ref{cross}).

We made ten Monte Carlo simulation runs, producing about 2500-3000 events in each of them, to simulate the statistics of the 2000 events in \cite{nakatwo}. We can make a statistical analysis of these ten different sets of results. The relative error is given by

\begin{align}
{\epsilon^{\,\prime}}^{2}_{r}=\frac{(N^{\,\prime})^{-1}\displaystyle{\sum_{i=1}^{N^{\,\prime}}g^{2}_{i}}-\Big[(N^{\,\prime})^{-1}\displaystyle{\sum_{i=1}^{N^{\,\prime}}g_{i}\Big]^{2}}}{\Big[(N^{\,\prime})^{-1}\displaystyle{\sum_{i=1}^{N^{\,\prime}}g_{i}\Big]^{2}}}\label{error}
\end{align}
where $N^{\prime}=10$ and now $g_{i}$ stands for the number of events falling inside a certain bin in each of the $i=1,...,N^{\prime}$ samples.

\begin{figure}
\centering
\includegraphics[width=0.55\textwidth]{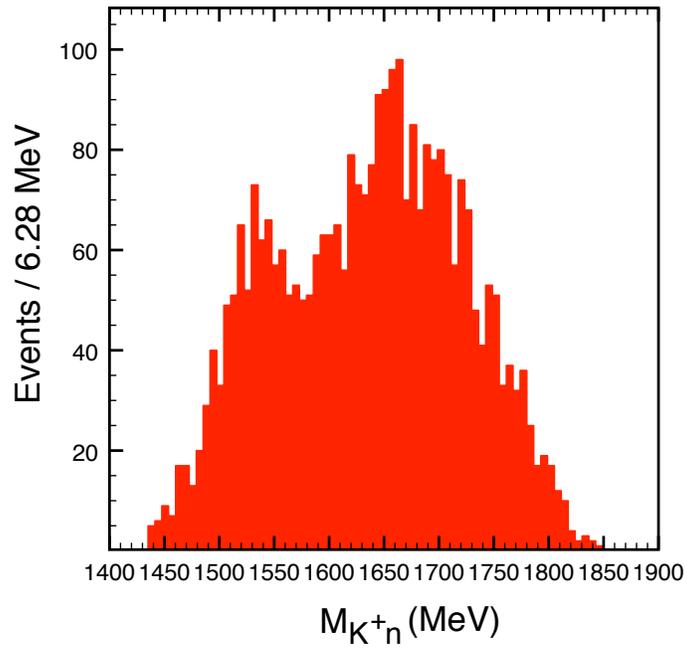}
\caption{(Color online) $M_{K^{+}n}$ distribution calculated with the MMSA prescription and the same cuts as those done in \cite{nakatwo} for 3098 events.}\label{1}
\end{figure}

\begin{figure}
\centering
\includegraphics[width=0.55\textwidth]{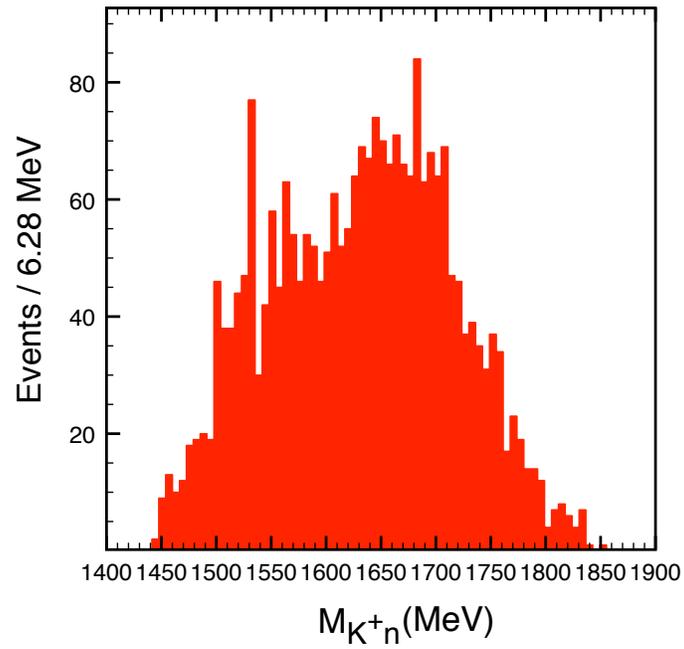}
\caption{(Color online) $M_{K^{+}n}$ distribution obtained with the MMSA prescription and the same cuts as those done in \cite{nakatwo} for 2523 events and a different seed than for  Fig. \ref{1} for the simulation run.}\label{2}
\end{figure}

\begin{figure}
\centering
\includegraphics[width=0.55\textwidth]{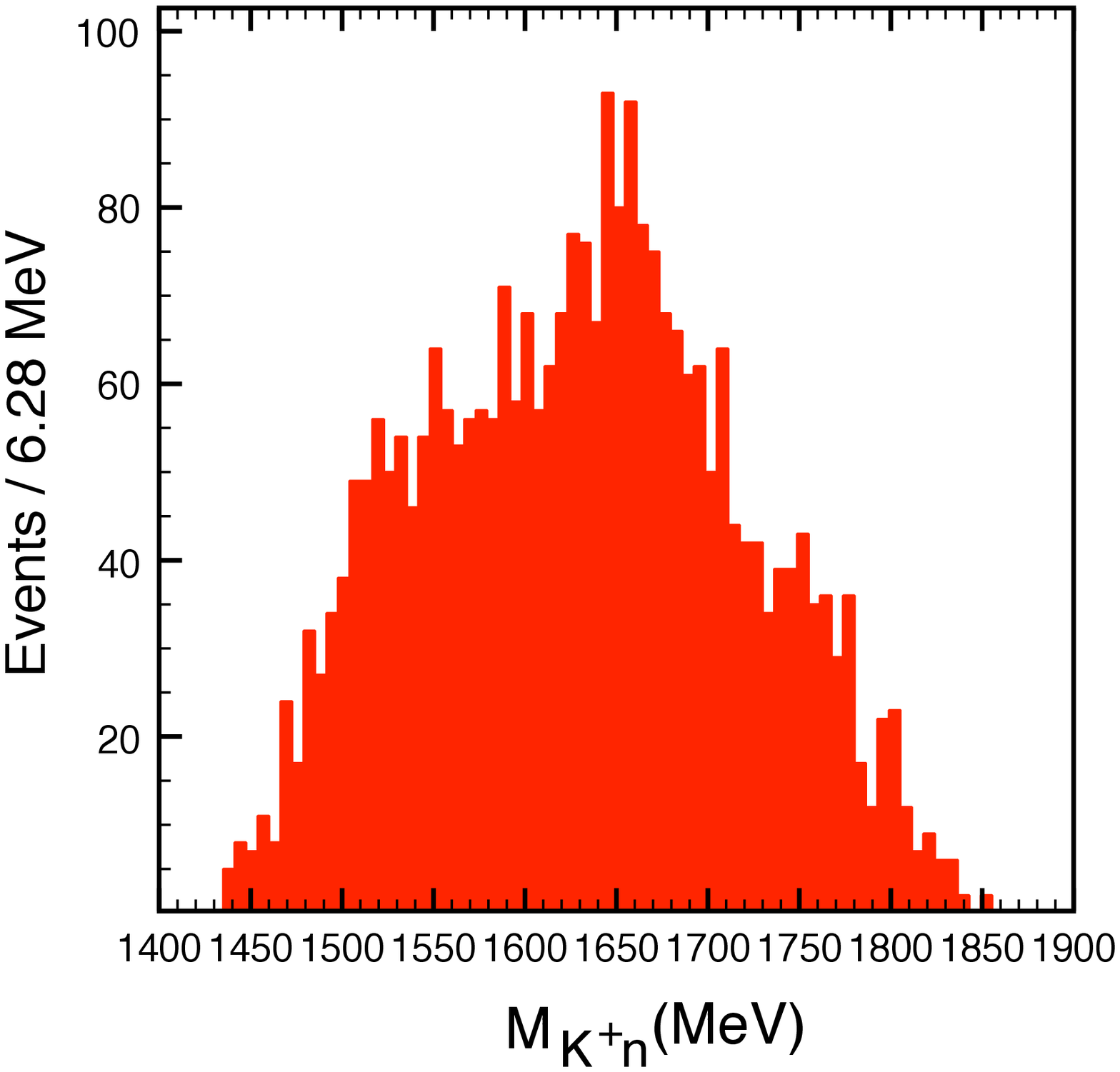}
\caption{(Color online) $M_{K^{+}n}$ distribution calculated with the MMSA prescription and the same cuts as those done in \cite{nakatwo} for 2627 events and a different seed than for  Figs. \ref{1}, \ref{2} for the simulation run.}\label{3}
\end{figure}

\begin{figure}
\centering
\includegraphics[width=0.55\textwidth]{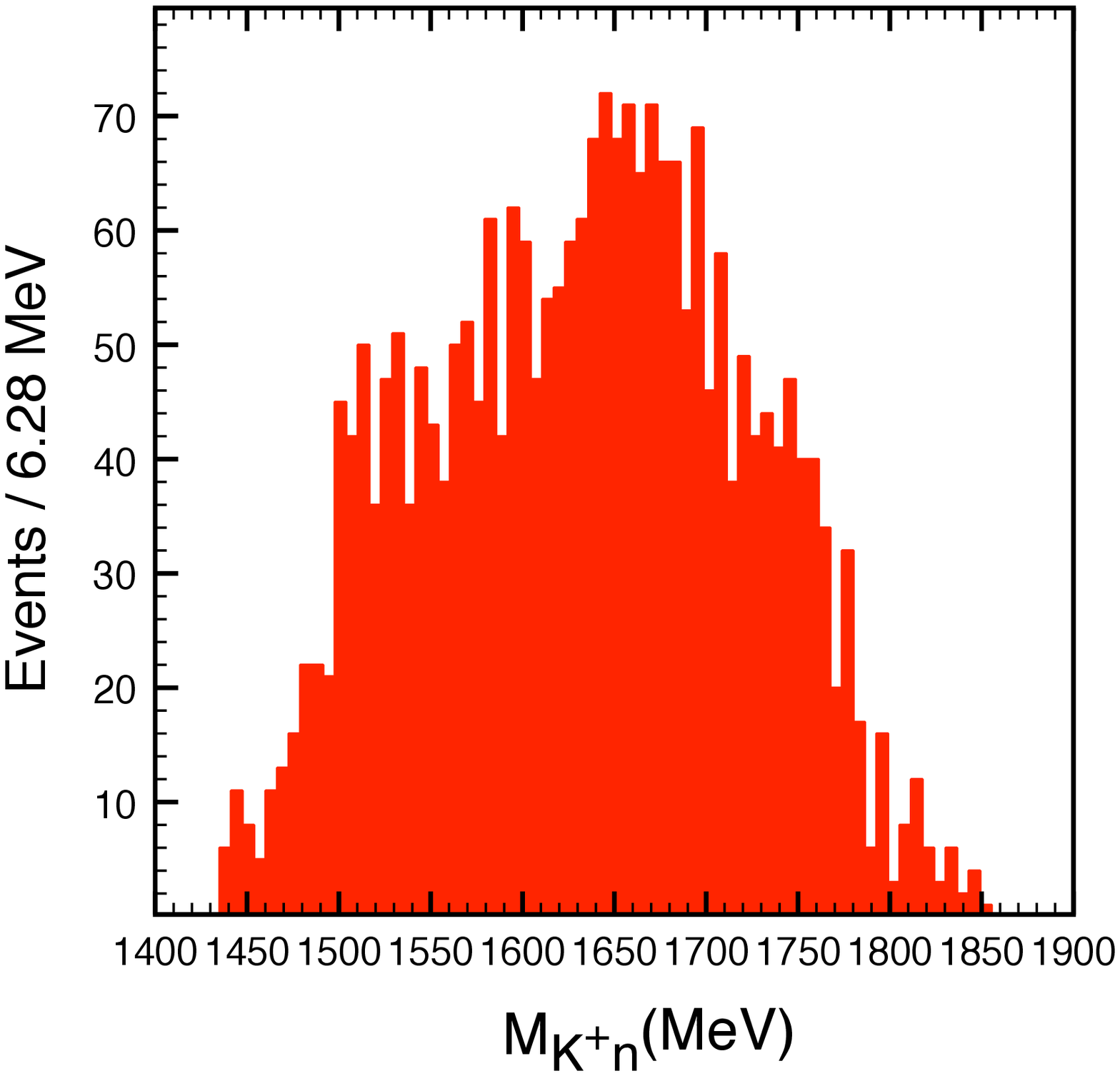}
\caption{(Color online) $M_{K^{+}n}$ distribution calculated with the MMSA prescription and the same cuts as those done in \cite{nakatwo} for 2502 events and a different seed than for  Figs. \ref{1}, \ref{2}, \ref{3} for the simulation run.}\label{4}
\end{figure}

We show in Figs. \ref{1}-\ref{4} four of the ten sets obtained with the simulation. We also take a different bin, 6.28 MeV, to facilitate the comparison with the data of \cite{nakatwo}. The error obtained with Eq. (\ref{error}) is of the order of 20 \% for the bins around $M_{K^{+}n}=1530$ MeV. The runs required 1210 million Monte  Carlo points.  We can see that in one of the runs we obtain a statistical fluctuation of about 20 counts over the peak at 1530 MeV of Fig. \ref{pentamore_Lambda}, indicating that, with these errors, statistical fluctuations of about the magnitude of the $\Theta^{+}$ peak seen in \cite{nakatwo} are possible. These errors in the simulation are similar to those of Figs. \ref{penta}, \ref{pentaLamb} using the Monte Carlo integral method, where similar statistical fluctuations were also observed.  It is also interesting to note that, like in the integration method, with the simulation we also observe clear peaks around 1530 MeV in about $1/3$ of the runs, and accumulation of strength in that region in practically all of them.

\begin{table}[h!]
\centering
\begin{tabular}{ccc}
\hline
Figure &  ``exact'' reference& $\chi^{2}/d.o.f$\\
\hline
\hline
Fig. \ref{penta}&Fig. \ref{pentamore}  & 1.50\\
Fig. \ref{pentades} & High statistic run with the same cuts &2.33\\
Fig. \ref{pentaLamb} & Fig. \ref{pentamore_Lambda}&1.65$^{\,a}$\\
Fig. \ref{1}  & Averaged ten Von Neumann runs&0.67\\
Fig. \ref{2}  & Idem& 1.81\\
Fig. \ref{3}  &Idem&  0.56\\
Fig. \ref{4}   & Idem& 1.00\\
\hline
\end{tabular}
\caption{Values of $\chi^{2}/d.o.f$ for different figures in the paper. $^{a}$Three points of the tail of the $M_{K^{+}n}$ distribution, around 1800-1850 MeV, of no relevance to the ``$\Theta^{+}$'' peak have been omitted in this evaluation of $\chi^{2}/d.o.f$.}\label{t1}
\end{table}

We also can do another test of the statistical significance of our runs. For this we can take advantage of the fact that we know the ``exact" curve from our large statistic integration, Fig. \ref{pentamore} for the $\phi$ production model and Fig. \ref{pentamore_Lambda} for the full model, or the average of the ten Von Neumann runs, which is already practically equal to Fig. \ref{pentamore_Lambda}. Then we can take the standard $\chi^{2}$ function of any run with respect to the exact curve.  The $\chi^{2}$ is defined as

\begin{align}
 \chi^2= \sum_{i}\frac{(f_i -f_{i}^{exact})^2}{\tilde{\sigma}^2}
\end{align}
where for $\tilde{\sigma}$ we take $\sqrt{f_i}$.  All the curves are normalized to the number of  events in the LEPS experiment, 1967 events for the $K^{+}n$ invariant mass.
In Table \ref{t1} we show the reduced $\chi^2$ (we have divided the $\chi^{2}$ by the number of degrees of freedom (d.o.f) which corresponds to the number of bins used) for the different figures that we have used in the discussion. 

We can see that the $\chi^{2}/d.o.f$  is between 0.56 and 1.81 for the Von Neumann runs.  The statistical significance of the results of Fig. \ref{penta}, and particularly, Fig. \ref{pentaLamb}, which is done with the full model, is similar to that of the Von Neumann runs. The $\chi^{2}/d.o.f$ for Fig. \ref{pentades} is a bit bigger than in the other cases. This run was obtained using the same seed as in Fig. \ref{penta} (equivalent to having the same experiment) and making a different cut. This figure is done with $\phi$ production alone and not the full model, so the results are only indicative. One can make a run with more points to have a smaller $\chi^{2}/d.o.f$ but the results do not change qualitatively. However, Fig. \ref{pentades} with the same seed as Fig. \ref{penta}, a similar number of points, and a different cut is more significative because it gives a hint of what might happen in the experiment, having a different shape than Fig. \ref{penta} and producing larger fluctuations. We do not have the experimental data for this cut and hence one cannot compare like in the other cases.

\begin{figure}
\centering
\includegraphics[width=0.7\textwidth]{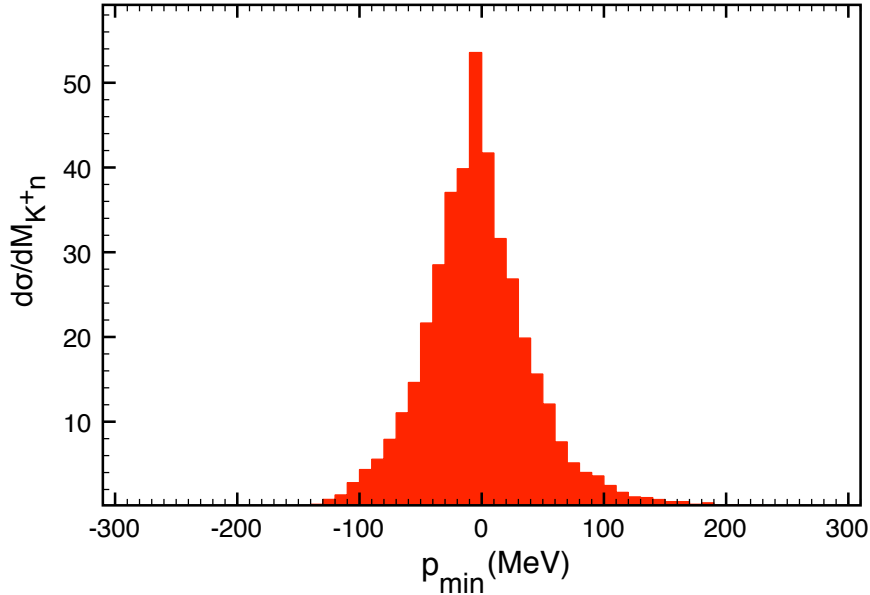}
\caption{(Color online) $p_{min}$ distribution.}\label{pmin}
\end{figure}

\begin{figure}
\centering
\includegraphics[width=0.7\textwidth]{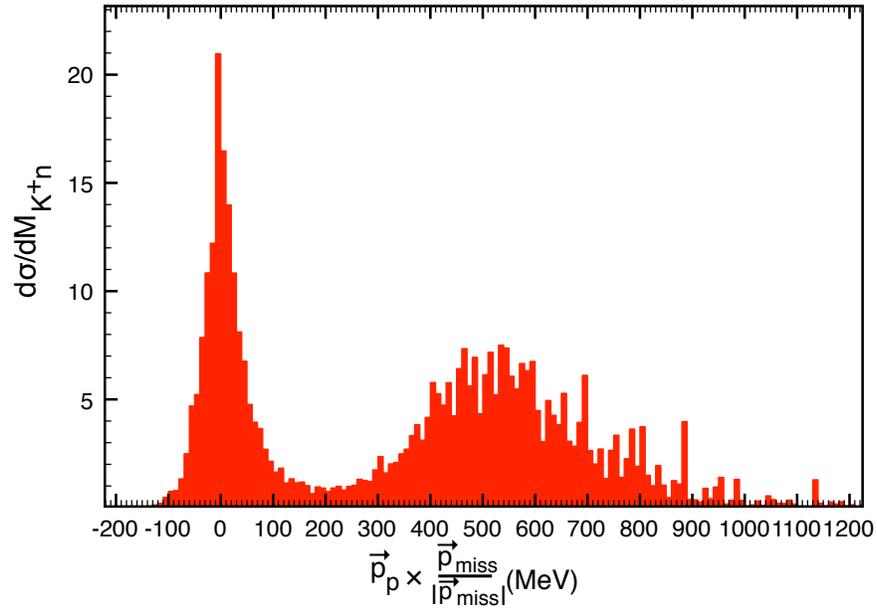}
\caption{(Color online) Distribution for the real proton momentum  component along the $\vec{p}_{miss}$ direction.}\label{pl}
\end{figure}

\begin{figure}
\centering
\includegraphics[width=0.7\textwidth]{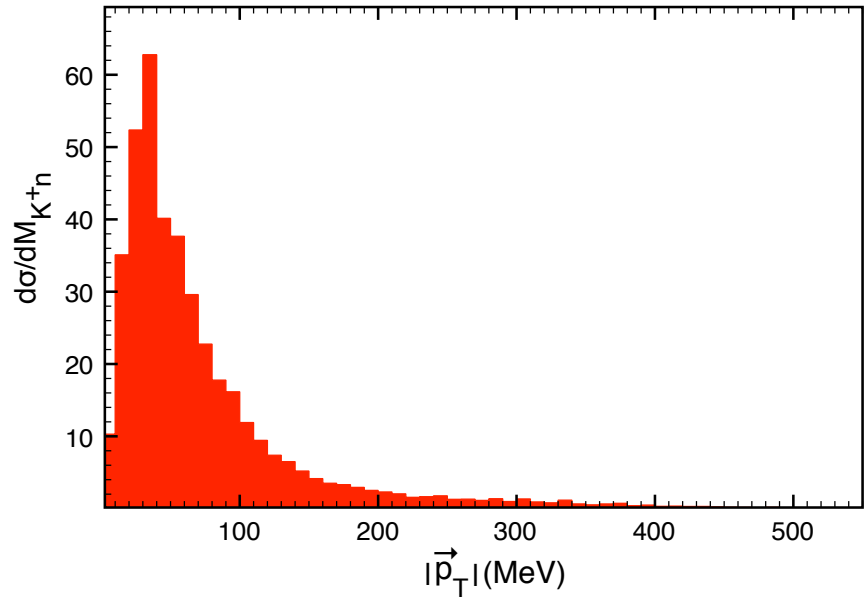}
\caption{(Color online) Distribution for the modulus of the real transverse proton momentum component, $|\vec{p}_{T}|$.}\label{pt}
\end{figure}

An interesting test that we can do to compare with the LEPS results is the distribution of $p_{min}$.  This is done in Fig. \ref{pmin} using the full model, i.e, including $\phi$ and $\Lambda(1520)$ production. We can see that we get one sharp peak around zero extending up to about -100 MeV to 100 MeV. The results are remarkably similar to those obtained in the LEPS experiment (see Fig. 4 of \cite{nakatwo}). Now it is equally interesting to make a plot for the distribution of proton momenta in the real case. In Fig. \ref{pl} we show the real distribution of the proton momentum component along the direction of $\vec{p}_{miss}$. We find a peak similar to the one of $p_{min}$, and a second broad peak, around 500 MeV, corresponding to protons which in the CM frame of the $np$ final state go in the same direction of $\vec{p}_{miss}$ rather than opposite to it, which is the situation that leads to $p_{min}$.  In the LEPS analysis these latter events would  be given a momentum $\vec{p}_{min}$ and no transverse component to the direction of $\vec{p}_{miss}$. This transverse component is also interesting to investigate. In Fig. \ref{pt} we show the distribution for the modulus of the transverse proton momentum component, $\vec{p}_{T}$. As we can see, the distribution peaks around 50 MeV and stretches up to about 200 MeV. 

It is easy to identify the sharp narrow peak of Fig. \ref{pl} with events where the proton is a spectator ($\phi$ production on the neutron), while the broad peak would come from events of $\Lambda(1520)$ or $\phi$ production on the proton, where the neutron would be a spectator.

\begin{figure}
\centering
\includegraphics[width=0.5\textwidth]{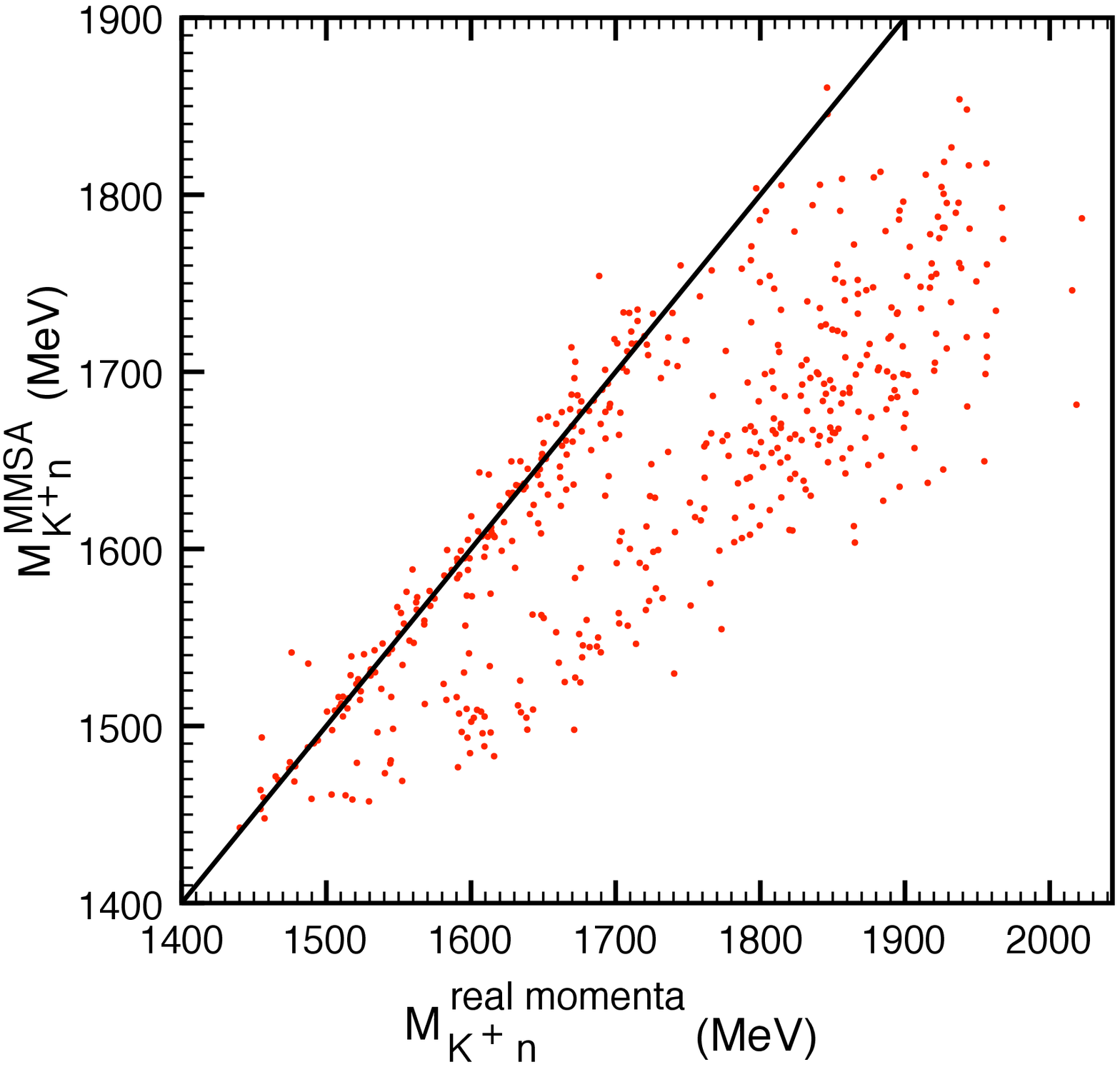}
\caption{(Color online) $M_{K^{+}n}$ calculated using the MMSA prescription versus $M_{K^{+}n}$ obtained with the real momentum for the nucleons and the full model, i.e., $\phi$ production on the nucleons and $\Lambda(1520)$ production on the proton.}\label{Minvnaka_real}
\end{figure}

\begin{figure}
\centering
\includegraphics[width=0.5\textwidth]{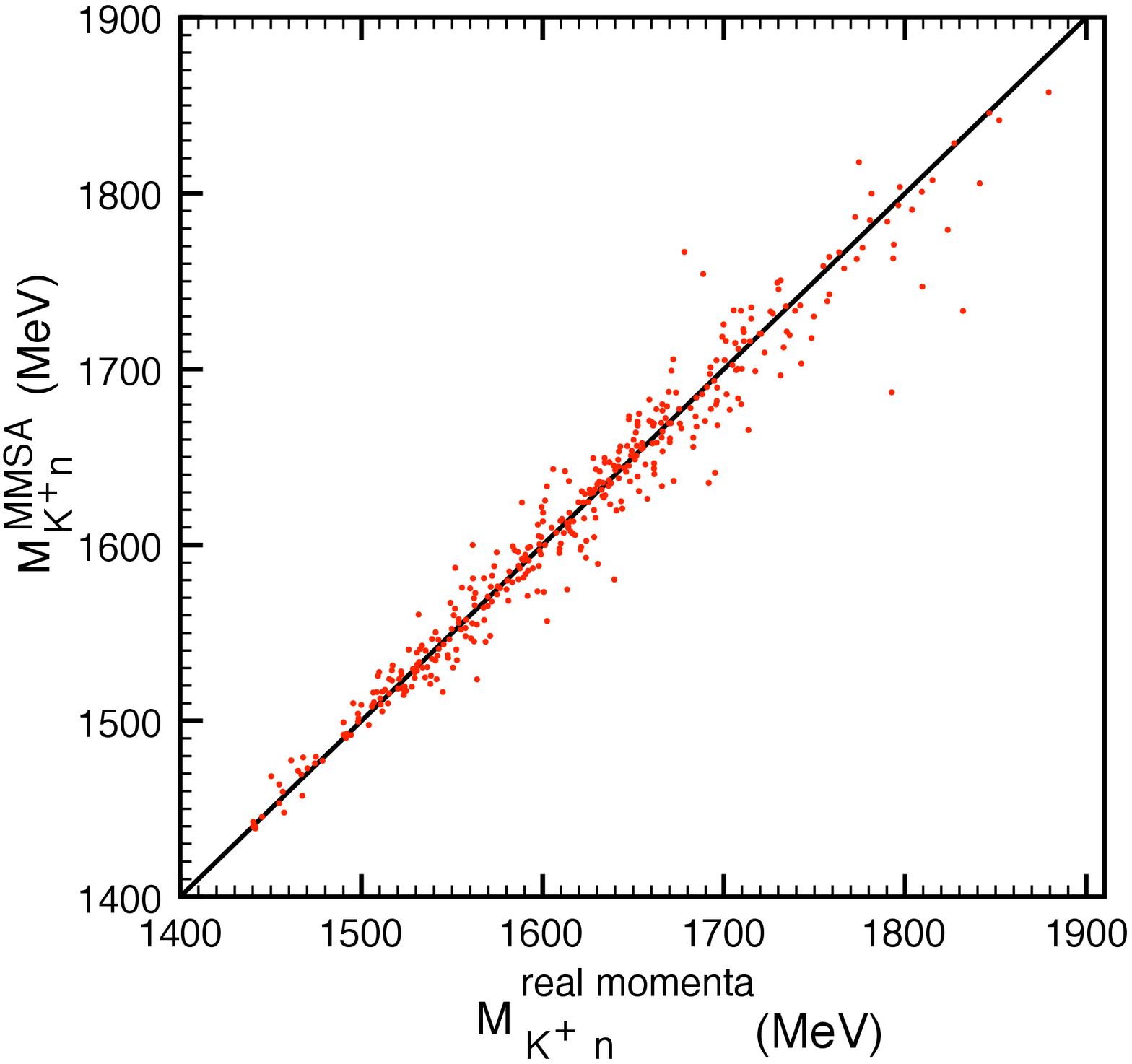}
\caption{(Color online) $M_{K^{+}n}$ calculated using the MMSA prescription versus $M_{K^{+}n}$ obtained with the real momentum for the nucleons omitting the double scattering, $\phi$ production on the proton and $\Lambda(1520)$ production on the proton.}\label{Minvnaka_real_singleterm}
\end{figure}

A complement to the further discussion can be seen in Fig. \ref{Minvnaka_real}, where we show a two dimensional plot for the correlation between the $K^{+}n$ invariant mass evaluated with the MMSA prescription and the real one. The events for which the two magnitudes are evaluated have been generated using the Von Neumann rejection method. We observe a relatively large dispersion of the points with respect to a diagonal line where the two invariant masses would be equal. Actually, we see two blocks of points, one around the diagonal and another one to the right of the diagonal at large $K^{+}n$ invariant masses. It is easy to identify the first block with the mechanism where the proton is a spectator while the second block corresponds to events where the neutron is the spectator. In order to visualize this assertion, we run again the simulation removing the $\Lambda(1520)$ and $\phi$ production on the proton and the double scattering. We show the results in Fig. \ref{Minvnaka_real_singleterm}. As we can see, now there is a very good correlation between the two invariant masses and the MMSA prescription should be a good tool to analyze a mechanism of this type. Of course, in the actual reaction one has both contributions from $\gamma$ scattering with the proton and the neutron and the realistic situation of the correlation between the two invariant masses is the one given in Fig. \ref{Minvnaka_real}.

The simulation runs, which require a large number of Monte Carlo points to get a few thousand of events, are suited to the actual experiment, and one should admit the same errors in the experiment of \cite{nakatwo} as we have found here. Also one should be ready to admit the possibility to have in the experiment the same statistical fluctuations over the peak of the large statistic distribution generated with the MMSA prescription.

There is one further element in favor of the former argumentation of the statistical fluctuations. In Fig. 10(a) of \cite{nakatwo} we see in the spectrum of the $M_{K^- p}$ a peak of about 20 events per bin around 1650 MeV. This peak is assumed in \cite{nakatwo} to be a statistical fluctuation, since in Fig. 16 of \cite{nakatwo}, making a cut $M_{K^+ K^-}>$ 1500 MeV, the peak disappears. Then one should note that the peak in $M_{K^+ n}$ seen at LEPS around 1526 MeV, when subtracted from the ``real" large statistics spectrum of our Fig. \ref{pentamore_Lambda}, corresponds also to about 20 events per bin ( see Fig. 12 of \cite{nakatwo}). The justification of the peak in the $M_{K^- p}$ spectrum as a statistical fluctuation has as a consequence the acceptance that the peak in the $M_{K^+ n}$ can equally be a statistical fluctuation. The large statistics calculation done here is important in this respect because it establishes the starting point from where to count the fluctuation. Since one already starts from a ``resonant like" structure around 1530 MeV, the diversion of the data of LEPS over this realistic spectrum is smaller than one might assume by inspection of Fig. 12 of \cite{nakatwo} alone.

\section{Conclusions}
We have made a simulation of the $\gamma d\to K^{+}K^{-}np$ reaction using a theoretical model that accounts for $\phi$  and $\Lambda(1520)$ production, the two visible resonance structures in the reaction of LEPS \cite{nakatwo}. The model takes into account rescattering of one of the kaons with the second nucleon in the deuteron. 

A Monte Carlo simulation of the reaction was done and distributions of $K^{+}n$ and $K^{-}p$ invariant masses were reconstructed. One of the elements in the LEPS analysis was the consideration of a $p_{min}$ of the nucleons, by means of which one could minimize the effect of Fermi motion of the nucleons in the deuteron and make an educated guess for the final, undetected neutron or proton momentum (MMSA prescription). We could see that the procedure allows to recognize a resonance if it is present in the theoretical amplitude. However, we also saw that the MMSA procedure, together with the cuts to eliminate the $\phi$ contribution, lead to a ``resonant like" structure around 1530 MeV in the $K^{+}n$ mass distribution, which we could show with a large statistics run. Since the peak seen at LEPS is narrower than this ``resonant like" structure, we made a thorough statistical analysis of the process to see if a statistical fluctuation could be responsible for it. For this purpose we made several simulation runs using the Von Neumann rejection method to generate events weighted by their probability of production, much as a Quantum Mechanical experiment does. The statistical study of these runs with a similar number of events as in the experiment \cite{nakatwo} provided the intrinsic error. These runs showed that a fluctuation over the peak generated by the MMSA prescription and the LEPS cuts around 1530 MeV was possible, with a similar size as the peak seen in \cite{nakatwo}.  On the other hand, in the  $M_{K^- p}$ spectrum of \cite{nakatwo} there is a peak around 1650 MeV which was assumed to be a statistical fluctuation since it disappeared making a slightly different $\phi$ cut. Then, using the  spectrum obtained with high statistics Monte Carlo runs we could see that the peak in $M_{K^+ n}$ in \cite{nakatwo} over this realistic spectrum has the same strength as the peak in $M_{K^- p}$, which was assumed to be a fluctuation in \cite{nakatwo}. The parallel study done at LEPS of the $M_{K^- p}$ spectrum has thus served to set a scale for the statistical fluctuations in the LEPS experiment, in such a way that one can accept the $M_{K^+ n}$ peak as a statistical fluctuation like the one admitted in \cite{nakatwo} for the peak in the $M_{K^- p}$ spectrum which has the same strength.

Another point worth making is that we could see in our Monte Carlo runs that for a same seed (equivalent to having a unique experiment) and different $\phi$ cuts, the $M_{K^{+}n}$ spectra changes appreciably, to the point that for some cuts the peak around 1530 MeV looks more prominent than for others. This means that, in principle, it is possible to convert an original fluctuation into one more pronounced by suitable changes in the cuts, which should be considered as a warning when analyzing this type of experiments. In fact, as mentioned above, the use of a $\phi$ cut $M_{K^{+}K^{-}}>$ 1050 MeV does not lead to a peak around $M_{K^{-}p}\sim 1650$ MeV in the experiment of \cite{nakatwo}, while the main $\phi$ cut used in \cite{nakatwo} gives rise to a peak at this value of the $M_{K^{-}p}$ invariant mass.

\section*{Acknowledgments}
This work is partly supported by the DGICYT contract  FIS2006-03438,
the Generalitat Valenciana in the program Prometeo and 
the EU Integrated Infrastructure Initiative Hadron Physics
Project  under Grant Agreement n.227431. One of the authors (A. M. T) thanks the support from a FPU grant
of the Ministerio de Educaci\'on.

\end{document}